\documentclass[sigconf]{acmart}

\usepackage{amsmath,amssymb,amsfonts}
\usepackage{algorithmic}
\usepackage{graphicx}
\usepackage{subfigure}
\usepackage{textcomp}
\usepackage{xcolor}
\usepackage{booktabs} 
\usepackage{makecell}
\usepackage{multirow}

\usepackage[linesnumbered,ruled,vlined]{algorithm2e} 
\usepackage{algorithmic}
\usepackage{fancyhdr}
\AtBeginDocument{%
  \providecommand\BibTeX{{%
    \normalfont B\kern-0.5em{\scshape i\kern-0.25em b}\kern-0.8em\TeX}}}

\renewcommand\footnotetextcopyrightpermission[1]{}

\begin{document}

\title{Edge-Assisted On-Device Model Update for Video Analytics in Adverse Environments}

\author{Yuxin Kong}
\affiliation{%
  \institution{Huazhong University of Science and Technology, Wuhan, China}
  \country{}
}
\email{yxkong@hust.edu.cn}

\author{Peng Yang}
\authornote{Corresponding author.}
\affiliation{%
  \institution{Huazhong University of Science and Technology, Wuhan, China}
  \country{}
}
\email{yangpeng@hust.edu.cn}

\author{Yan Cheng}
\affiliation{%
  \institution{Huazhong University of Science and Technology, Wuhan, China}
  \country{}
}
\email{y_cheng@hust.edu.cn}


\begin{abstract}

While large deep neural networks excel at general video analytics tasks, the significant demand on computing capacity makes them infeasible for real-time inference on resource-constrained end cameras. In this paper, we propose an edge-assisted framework that continuously updates the lightweight model deployed on the end cameras to achieve accurate predictions in adverse environments. This framework consists of three modules, namely, a key frame extractor, a trigger controller, and a retraining manager. The low-cost key frame extractor obtains frames that can best represent the current environment. Those frames are then transmitted and buffered as the retraining data for model update at the edge server. Once the trigger controller detects a significant accuracy drop in the selected frames, the retraining manager outputs the optimal retraining configuration balancing the accuracy and time cost. We prototype our system on two end devices of different computing capacities with one edge server. The results demonstrate that our approach significantly improves accuracy across all tested adverse environment scenarios (up to 24\%) and reduces more than 50\% of the retraining time compared to existing benchmarks.

\end{abstract}





\maketitle

\section{Introduction}

Empowered by the advancement of deep neural networks (DNNs) and the increasing prevalence of cameras, video analytics has become an indispensable component in various applications, such as traffic monitoring \cite{survey,chame,drl,spatula}, autonomous driving \cite{a7,a1,jerry,yan}, intelligent industry \cite{a3,scale}, etc. While advanced DNNs can generate accurate inference results for various challenging analytics tasks \cite{vod, flat}, the sophisticated network structures and substantial parameters result in a significant computation burden that renders such models impractical for real-time analytics on resource-constrained end devices \cite{glimpse,scale2}, e.g., most of the off-the-shelf cameras.

\begin{figure}[t]
    
  \begin{minipage}[t]{0.485\linewidth}
      \centering
      \subfigure{
          \includegraphics[width=1\linewidth]{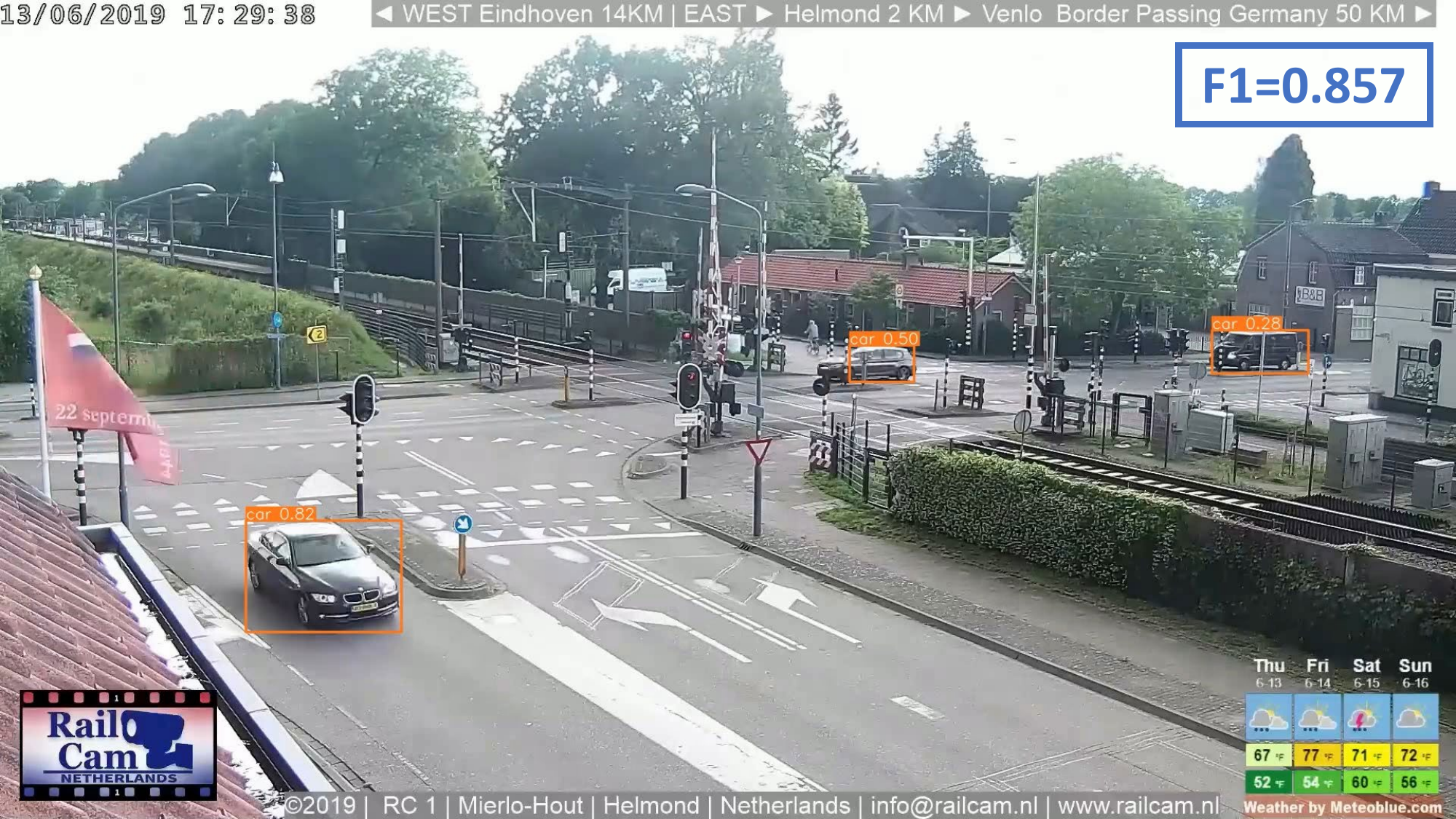}
      }
      \subfigure{
          \includegraphics[width=1\linewidth]{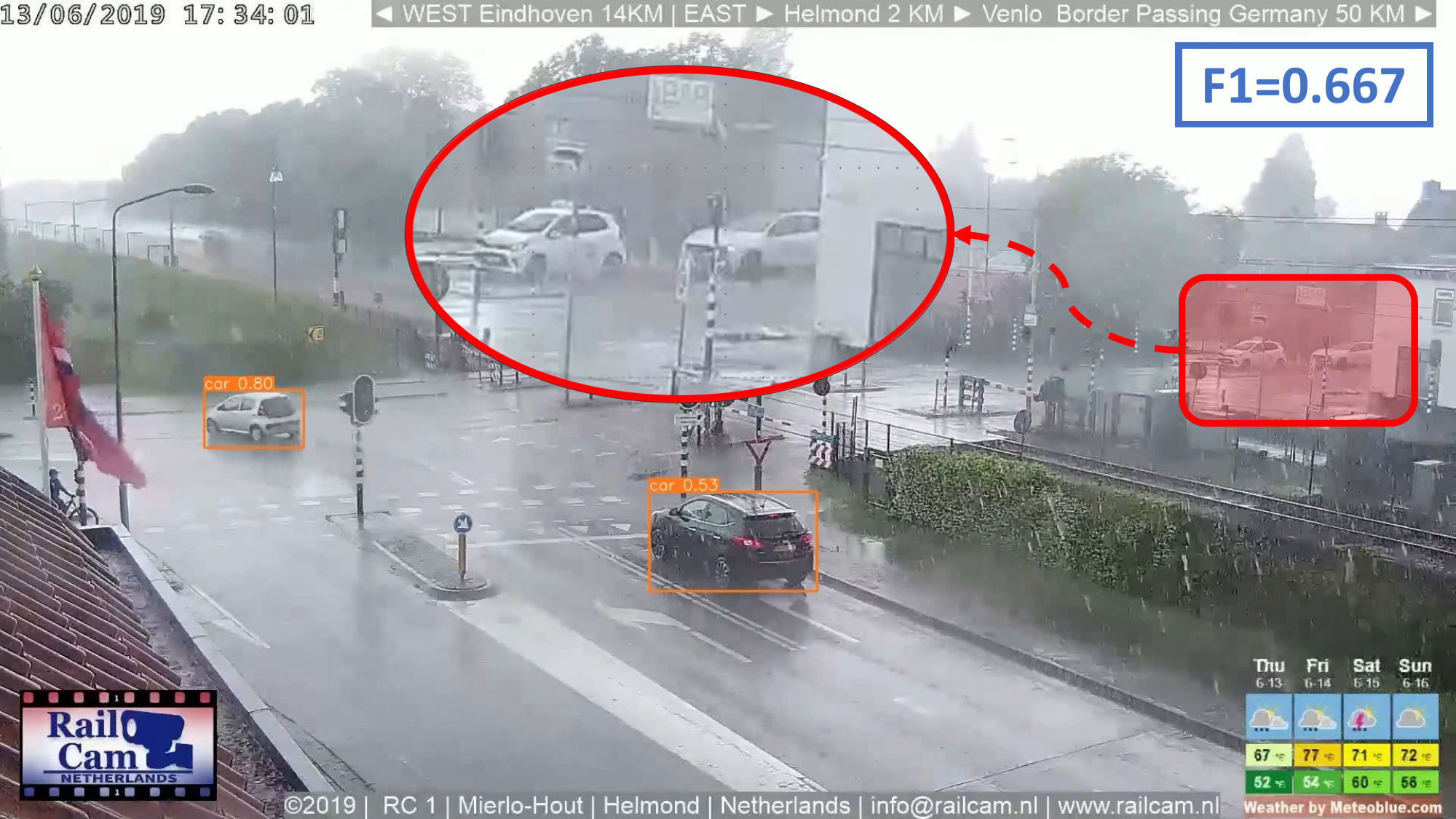}
      }
  \end{minipage}
  \begin{minipage}[t]{0.485\linewidth}
      \centering
      \subfigure{
          \includegraphics[width=1\linewidth]{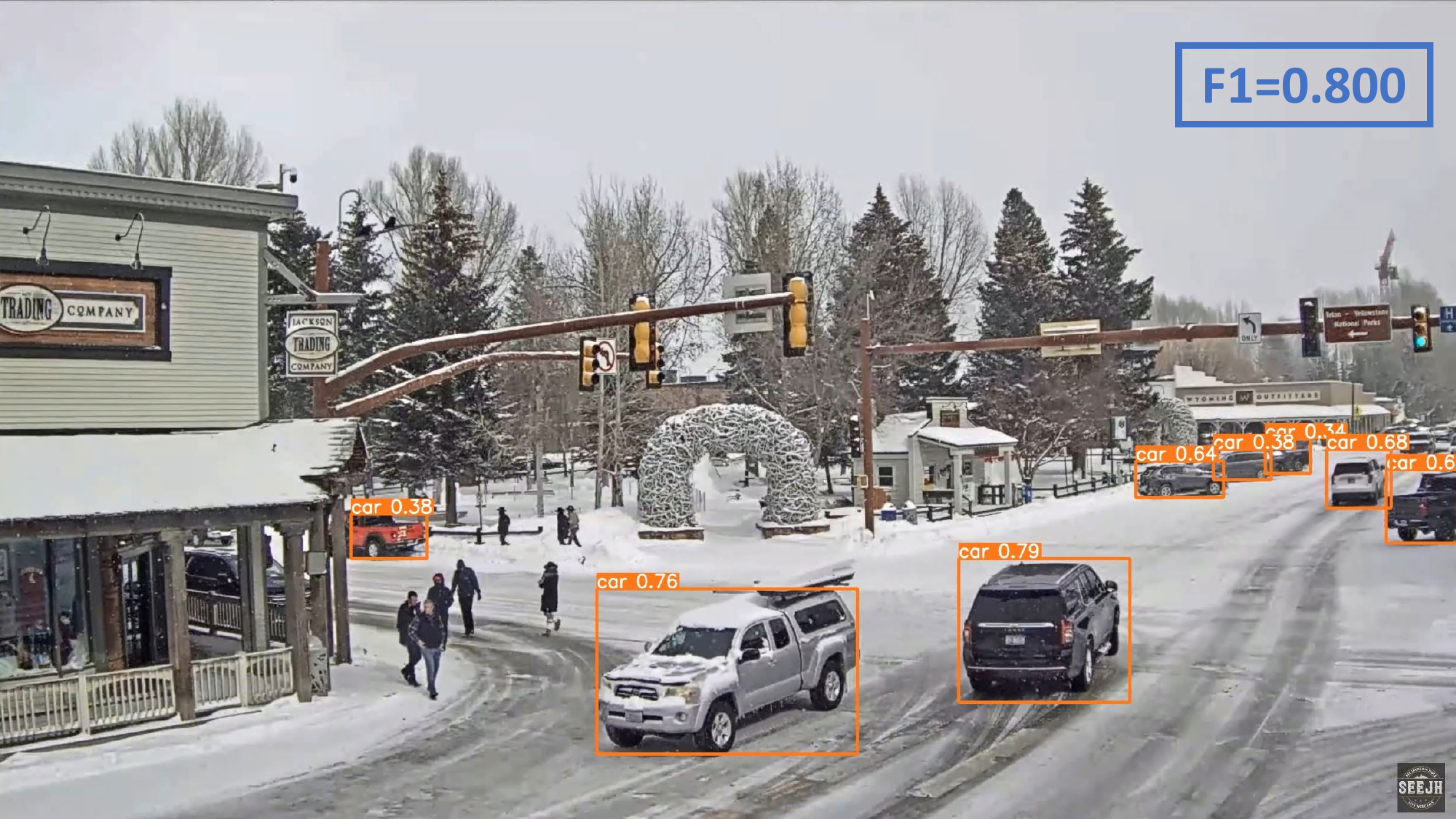}
      }
      \subfigure{
          \includegraphics[width=1\linewidth]{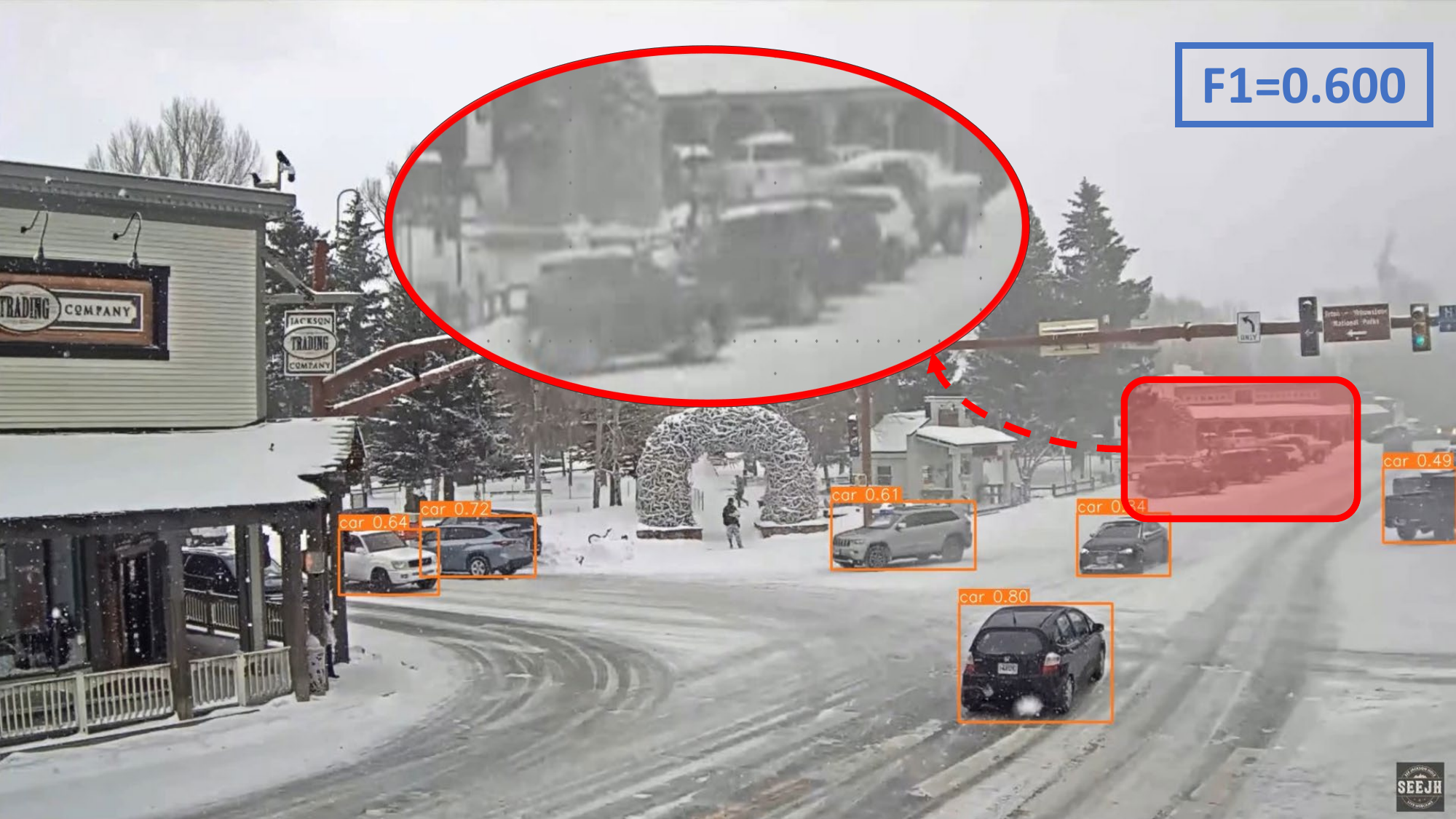}
      }
  \end{minipage}
  \caption{A comparison of lightweight Yolov5-n \cite{c1} performing object detection under normal environments (Row 1) and adverse environmental conditions (Row 2).}
  \label{bad}
\vspace{-5mm}
\end{figure}

One common approach to tackle this challenge is to employ carefully pruned or compressed models that can adapt to the restricted computing resources of end devices, which has been proven effective in well-conditioned environments or under customized scenes \cite{b1,b2,noscope,spe}. Nevertheless, it is difficult for lightweight models to achieve high-performance inference results in diverse scenarios due to the intrinsic simplified architecture and limited parameters, especially under challenging adverse environments, such as heavy rain and heavy snow. As shown in Fig. \ref{bad}, comparative results of captured video frames have been presented under normal and adverse environments. We observe that, despite similar objects being in nearly identical positions, they cannot be detected under adverse conditions (circled by the red line at Row 2), leading to a large accuracy drop of 20\%, whereas they can be accurately identified under favorable environments using the same lightweight model.

Multiple factors can cause adverse environments, such as poor lighting conditions and extreme weather \cite{aqua,yuan}. In these cases, the video imaging condition significantly deteriorates, manifesting in three distinct differences in the captured frames: increased noise, changed brightness, and blurred objects. All of them can impair the feature extraction of objects, making inference more difficult. To enhance the performance of lightweight models in non-ideal environments on end devices while maintaining real-time inference, continuous learning has emerged as a promising technique \cite{d1,recl,rilod}. This approach continuously retrains the lightweight model, or the student model on new data samples, leveraging the knowledge transferred from a more advanced DNN, which is referred to as the teacher model. The new data samples involve specific information relevant to the current environment, allowing the student model to specialize in the given scenario. By retraining the student model with new data samples, it can adapt to changes in the environment, therefore improving overall performance.

However, the retraining process is computationally intensive. It also requires groundtruth labels from a large teacher model, rendering it infeasible to be efficiently executed on resource-constrained end devices \cite{aws,acm1,acm2,reducto,elf}. To address this challenge, in this paper, we utilize an edge server to assist with retraining tasks. This is because edge computing can help to meet the rigorous demands of video analytics tasks due to its availability in the proximity of source video cameras \cite{a5,emerging,deep,dds,a2,review}. Besides, less transmission bandwidth consumption and lower response latency can be achieved, compared with cloud-based solutions \cite{a4,jcab,casva,distream}. Instead of offloading all the videos and tasks to the edge server, we aim to perform inference tasks on end cameras and conduct retraining tasks on edge servers, thereby leveraging the computing resources at both sides, as well as meeting bandwidth constraints.

In particular, we propose an edge-assisted real-time video analytics system that aims at enhancing the performance of the student model on end cameras under adverse environments. Once the environmental condition is detected to be changed, the student model on the end cameras is updated with the retrained weight parameters from the edge server. Nonetheless, implementing such a system is non-trivial as it faces three significant challenges. \emph{Firstly}, how to efficiently select data samples for better retraining while the size of data is subject to bandwidth constraints. \emph{Secondly}, how to trigger the retraining timely while avoiding unnecessary retraining process. \emph{Finally}, how to achieve optimal retraining configurations for better retrained model accuracy and faster response. To address those challenges, we first devise an efficient Key Frame Extractor which uses the inference results of the student model to obtain the frames that can best represent the current environments and removes temporal redundancy to meet the current bandwidth constraints. Then, an accuracy-based Trigger Controller is designed to adaptively trigger the retraining process. Moreover, we formulate an optimization problem that aims at selecting the best configurations for retraining tasks to strike a balance between accuracy and total latency. A heuristic algorithm is proposed in the Retraining Manager module to solve the optimization problem. Finally, we develop a test-bed consisting of two end cameras and an edge server to evaluate the performance of the proposed framework. Our main contribution can be summarized as follows.

\begin{itemize}
	\item We propose a key frame extraction method, which fully utilizes the student inference results to efficiently select the most informative frames for retraining. Temporal redundancy is also removed based on frame difference to meet the bandwidth constraints. This two-step key frame extractor is low-cost and hence can be implemented on end cameras.
	\item We design a mechanism to adaptively trigger the retraining task based on the accuracy drop of sampled frames, which effectively prevents unnecessary retraining that wastes computation resources on the edge server while maintaining satisfying inference performance.
	\item We devise a Retraining Manager module to identify the optimal retraining configurations, the resulting algorithm can achieve a balanced tradeoff between inference accuracy and total retraining time overhead.
  \item We prototype our system on two end devices and one edge server. Then, we validate our approach under three typical adverse environments with the task of object detection. The proposed solution is shown to have up to 24\% accuracy improvement and 50\% retraining time reduction.
\end{itemize}

The remainder of this paper is organized as follows. Section \ref{Motivation} describes the background and illustrates relevant experiment that motivates our work. Section \ref{System} presents the details of the system model design. Section \ref{Evaluation} demonstrates the evaluation results. Finally, we conclude our paper in Section \ref{Conclusion}.

\section{Background and Motivation}\label{Motivation}

Recent advances in computer vision have led to the development of state-of-the-art (SOTA) DNNs, which are designed with deeper network layers and wider feature maps, making them capable of covering a wide range of scenes and vision tasks \cite{temp,crossroi,dt}. However, the ever-increasing complexity of DNN architectures also increases the prediction latency and poses a significant challenge to delay-sensitive tasks on resource-constrained end devices \cite{effi, mobi}.

\begin{figure}[tbp]
  \centering
  \subfigure[Accuracy v.s. time curve of different models or methods.]{\includegraphics[height=3cm,width=4cm]{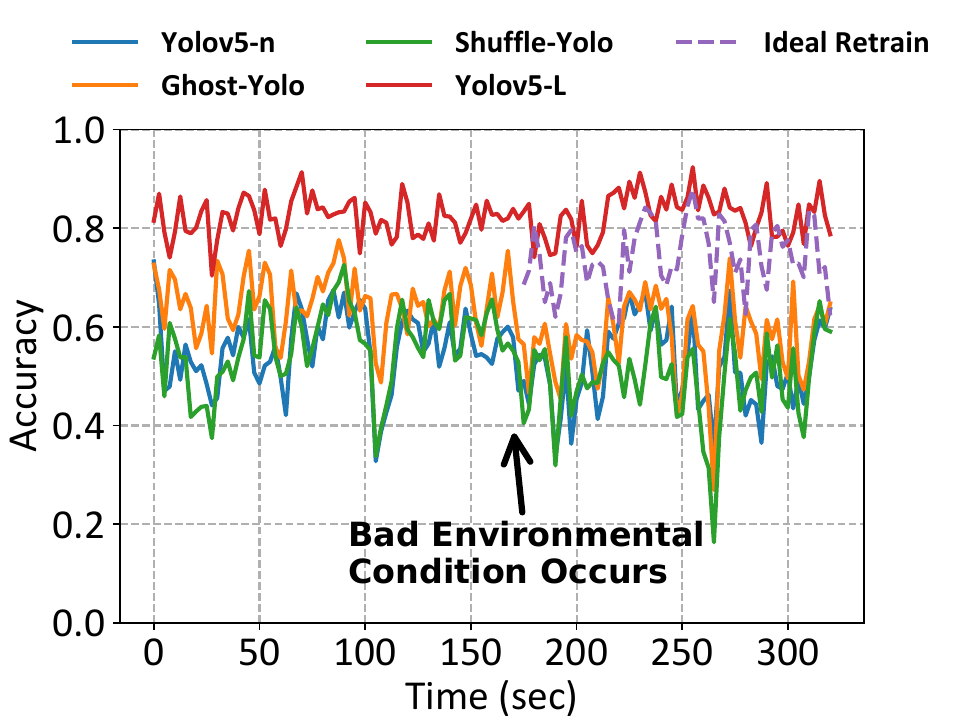}}
  \subfigure[Latency and average accuracy before and after adverse environments.]{\includegraphics[height=3cm,width=4cm]{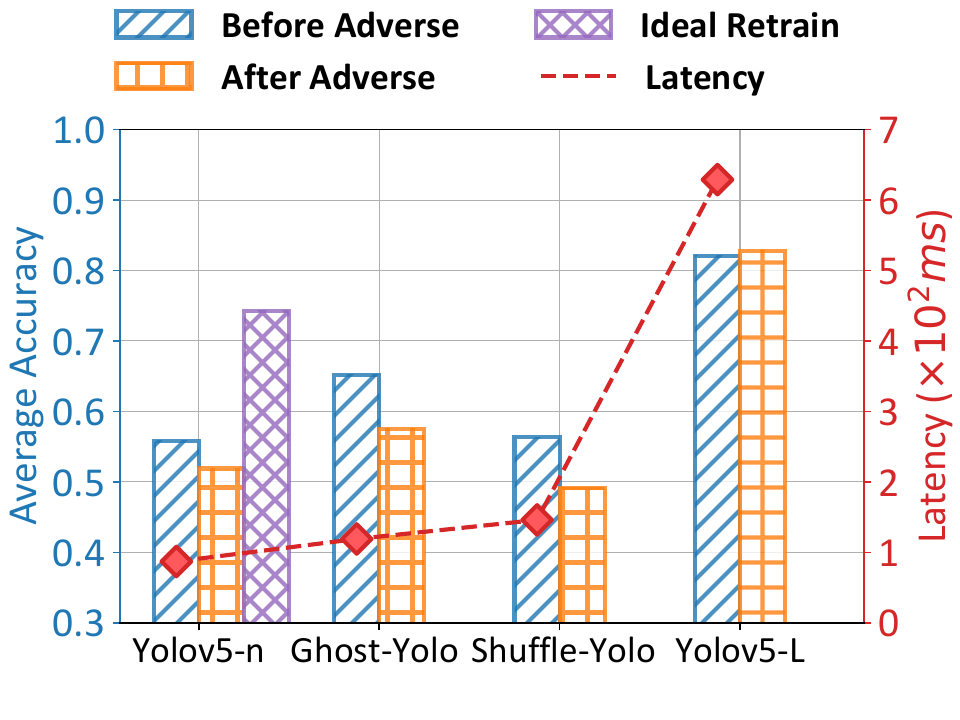}}
  \caption{Comparison of different methods during which adverse environments occur at $t=170s$.}
  \label{moti}

\end{figure}

\begin{figure*}[t]
  \centering
  \includegraphics[scale=0.47]{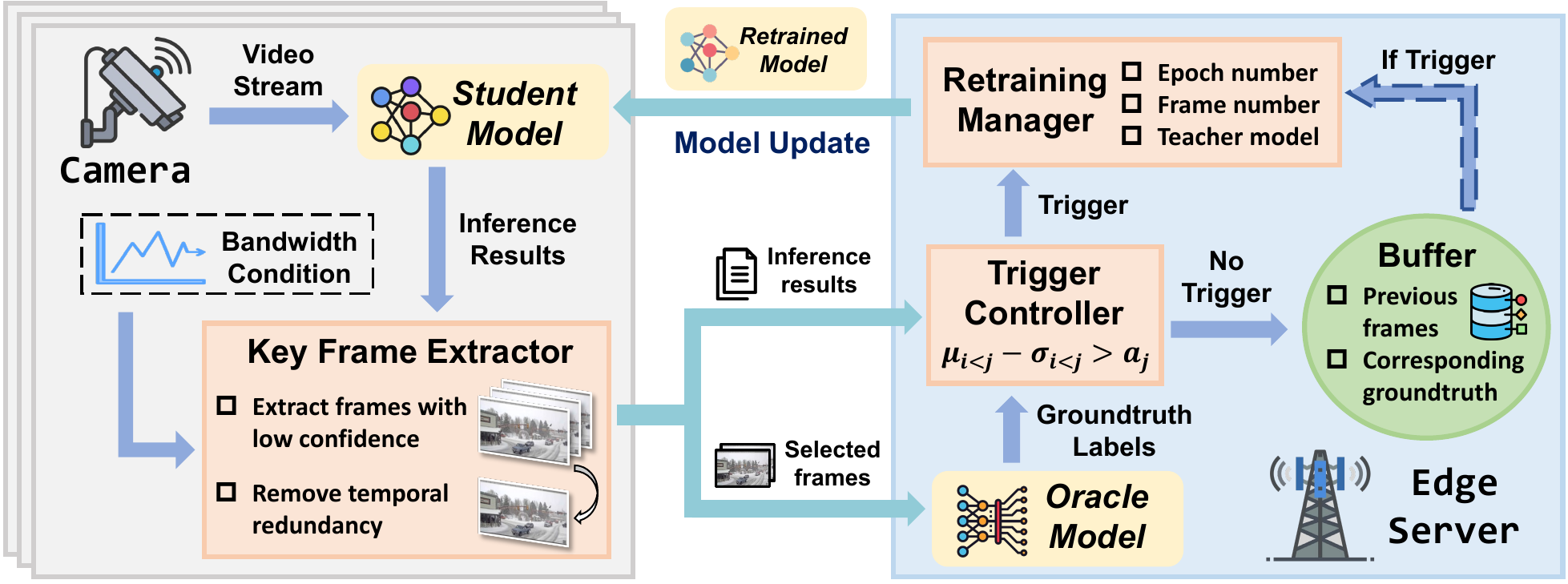}
  \caption{An overview of system design.}
  \label{framework}
\end{figure*}
In order to balance the tradeoff between accuracy and latency on resource-constrained end devices, many existing works have exploited the potential of lightweight models with simplified network architecture and reduced amount of parameters, such as GhostNet \cite{b1} and ShuffleNet \cite{b2}, which can deliver remarkable performance in either customized environments or favorable conditions. Nevertheless, although lightweight, the inherent structural characteristics preclude such models from achieving high performance in more challenging scenarios, such as those with dynamic object distributions or adverse environmental conditions. To tackle the first type of challenging scene, continuous learning-based methods have been proposed by many works \cite{Ekya,AMS,JIT}. Bhardwaj \emph{et al.} have illustrated that the varying object class distributions degrade the efficacy of lightweight models, while retraining is effective in enhancing accuracy under such circumstances \cite{Ekya}. 

We conduct the following experiments to investigate the ability of SOTA lightweight models in handling the second type of challenging scenarios, \emph{i.e.}, under adverse environments. We compare Yolov5 series models \cite{c1}, GhostNet \cite{b1}, and ShuffleNet \cite{b2}, respectively on NVIDIA Jetson Nano, which is the most widely used yet resource-constrained end devices. We replace the Yolo backbone with GhostNet and ShuffleNet backbones to take advantage of the faster feature extraction capabilities while maintaining the same Yolo detection head. Specifically, we use a 320s video from YouTube\footnote{https://www.youtube.com/watch?v=1EiC9bvVGnk}, during which heavy snow suddenly occurs after the $170-th$ second and persists for the rest of the time.  Additionally, we retrain Yolov5-n with the best retraining configurations (obtained via exhaustive searching) and we disregard any potential increase in latency caused by the retraining process to realize \emph{accuracy-ideal} retraining. In this way, we can investigate whether retraining can yield better performance in such scenarios. The F1-Score results over time are depicted in Fig. \ref{moti}(a), while Fig. \ref{moti}(b) presents a comparison of average accuracy before and after exposure to adverse environmental conditions, as well as the inference latency per frame. Our key observations are as follows.

\begin{itemize}
	\item Heavy DNN generates better results with higher latency. Yolov5-L keeps much more stable accuracy all the time and outperforms all lightweight models by a significant margin, even when faced with sudden environmental degradation. However, the heavy computation overhead results in almost six times more latency than other models, making it impractical for deployment on such resource-constrained devices.
	\item SOTA lightweight models cannot cope with bad environmental conditions. Even the two SOTA lightweight models exhibit a sharp drop in accuracy around $t=270s$ and suffer obvious average accuracy loss from the poor environment. 
	\item Retraining is promising to enhance performance. The retrained Yolov5-n can yield substantial accuracy benefits under ideal conditions, but it is not feasible to determine the optimal frames and epochs for each retraining instance as the prolonged retraining latency may render the retrained model obsolete for ever-changing video streams.
\end{itemize}

In summary, continuous retraining is an effective strategy to improve model performance on resource-constrained end devices when dealing with unfavorable environmental conditions. Moreover, it is essential to ensure that each retraining is configured properly to optimize overall accuracy and retraining latency.

\section{System Model}\label{System}

In this section, we present the design of our edge-assisted real-time video analytics system. We start by providing an overview of the system and then discuss its three key components respectively: Key Frame Extractor, Trigger Controller, and Retraining Manager.

\subsection{System Overview}
As shown in Fig. \ref{framework}, our video analytics system consists of modules on the end camera and the edge server.

\textbf{End camera side.} A lightweight student model is deployed on the end camera to perform real-time inference, with adaptive updates to accommodate the ever-changing environmental conditions. We develop a Key Frame Extractor that leverages the inference results of the student model to identify frames that can best represent the current environmental conditions within a given time window. Besides, we remove the temporal redundancy via frame differencing to meet the bandwidth requirements, as we offload the retraining task to the edge server. Then, these selected frames and the corresponding inference results of the student model are transmitted to the edge server for further analysis.

\textbf{Edge server side.} After transmitting the data of a window to the edge server, a sophisticated Oracle model generates groundtruth labels for these frames. A Trigger Controller is designed to make adaptive decisions of whether initiate retraining based on the empirical accuracy variation. If retraining is considered to be unnecessary, the received data is stored in the Buffer and used as training samples for the next retraining process. Otherwise, the Retraining Manager leverages the data of both the present window and buffered key frames to obtain a sub-optimal configuration for balancing accuracy and retraining time overhead. Besides, we incorporate the knowledge distillation technique into our retraining process, given its widely accepted role in enabling small models to acquire new complex knowledge quickly \cite{b7, obj200}. As the selection of different teacher models can impact both accuracy and time cost, we also regard it as a tunable parameter in Retraining Manager, in addition to the epoch number and frame amount. Finally, after the retraining process is done, we update the current model on the end camera with the retrained model to achieve more accurate inference results.

\subsection{Key Frame Extractor}

Performing retraining task demands substantial computational power and cannot be carried out on end cameras with limited resources which result in high latency \cite{videoedge,offload,a6}. To address this issue, we leverage an edge server to assist in retraining, which entails transmitting data samples from end camera to edge server.

\begin{algorithm}[t] 
	\caption{Key Frame Extraction Algorithm}
	\KwIn{$\mathcal F = \{f_1, f_2, ..., f_i, ...\}$: frame set of current window, $\beta $: filtering threshold, $B_{up}$: current uplink bandwidth. \\
            }
	\KwOut{
		$\mathcal {\widetilde{S}}$: Low confidence frame set, $\mathcal S$: List of selected frames for transmission.
            }
    {  

        \textbf{Step1: Extract Low-confidence Frames}

        \For{$f_{i} \ in \ \mathcal{F}$}
    {
        $L_{i}$ $\gets$ getLowConfRatio($f_{i}$)
        
        \If{$L_{i} > \beta $}
        {
            $\mathcal S.append(f_{i})$
        }    
    }

    $\mathcal {\widetilde{S}} = \mathcal S$ \quad$\triangleright$ the set of low confidence frames

        \textbf{Step2: Remove Temporal Redundancy}

        $C$ $\gets$ determine the maximum number of transmitted frames limited by the current available uplink bandwidth $B_{up}$
    
    \If{$| \mathcal S|  \leq  C$}
    {
      \Return $\mathcal S$
    }
    \Else
    {
      \For{$f_{i} \ in \ \mathcal{S}$}
    {
        $d_{i}$ $\gets$ getFrameDifference($f_{i}$)
    }
    
    $\mathcal {S^*}$ $\gets$ sort $\mathcal S$ by $d_{i}$ with a descending order

    $\mathcal {S}$ = $\mathcal {S^*}[:C]$ \quad$\triangleright$ select top-$C$ frames with the largest frame difference

    \Return $\mathcal S$ 
    }
    }
  
\end{algorithm}

Video frames can offer a plethora of information, but not all of them necessarily contribute to enhancing the performance of model retraining. The reasons are two-fold. First, during the training process, the model parameters are updated only when there are discrepancies between the predicted labels and the groundtruth labels. As a result, frames that the student model can accurately predict do not provide sufficient helpful information for retraining. Second, consecutive frames in the video sequence contain redundant information that does not offer extra gain to the training process. Besides, aggressively sending all of the frames to the edge server for selection is bandwidth inefficient. As a result, conducting key frame extraction directly on device is imperative.


We divide the video stream into windows of equal size and conduct key frame extraction within each window once the detection of the current window is completed. By leveraging the confidence score of student inference results, we can assess if the current student model can accurately predict the current frames, as lower confidence scores suggest higher inference difficulty, indicating possible adverse environments. Moreover, we adopt a frame differencing approach to reduce the temporal redundancy, which enables us to meet the stringent bandwidth requirements in unfavorable network conditions. The frame differencing method is a low-cost technique that allows for the detection of pixel-level changes in brightness, color, and texture within video sequences \cite{framediff}.

Algorithm 1 summarizes the process of our two-step Key Frame Extractor within a time window. \emph{Firstly}, it iterates over each frame in the window and calculates the low-confidence ratio, defined as the proportion of detected objects with low confidence. If the ratio exceeds the pre-defined threshold $\beta$, indicating that the majority of objects in the frame are difficult for prediction, this frame is then added to the selected list $\mathcal{S}$. \emph{Next}, we determine the maximum number of frames $C$ that can be transmitted under the current uplink bandwidth limit $B_{up}$. If the number of frames in $\mathcal{S}$ is less than $C$, we transmit the whole list directly to the edge server. Otherwise, we compute the frame difference for each frame in $\mathcal{S}$. Since the list $\mathcal{S}$ is only a subset of $\mathcal{F}$ after Step 1, the calculations can be cut down dramatically. As a result, we select top-$C$ frames with the largest frame difference as the most representative frames for offloading.

This low-cost yet effective approach allows us to identify the most informative frames while adhering to bandwidth requirements. In addition, along with the selected frames, the student inference results will also be transmitted to the edge server for further analysis. The data size of this supplementary information is negligible.

\subsection{Trigger Controller}

Previous works such as AMS \cite{AMS} and Ekya \cite{Ekya} adopt a fixed update interval. Nonetheless, a large interval may fail to respond timely to sudden changes in the current environment, while a low update interval may cause unnecessary frequent retraining that wastes computational resources on the edge server. Other study like JIT \cite{JIT} triggers retraining once detected the accuracy of the student model falls below a pre-defined threshold. However, a fixed threshold may not be suitable for all scenarios with varying levels of difficulty. Based on this idea, we devise a subtle trigger method to make the threshold adaptive, realizing both prompt response to environmental changes and efficient use of resources. 

After the selected frames and corresponding student inference results of a window are transmitted to the edge server, an advanced Oracle model generates groundtruth labels for these frames. The groundtruth labels are then used to retrain the model and evaluate the accuracy of those selected frames within the current window. By jointly taking into account the accuracy results of each received window, we develop an adaptive Trigger Controller.

For each retraining cycle, let $\mathcal{W} = \{W^{1}, \dots, W^{i}, \dots, W^{j}\}$ denote all the windows received by the edge server up to the current window $W^{j}$. Correspondingly, we define the average accuracy of these windows as $\mathbf{a} = \{a_{1}, \dots, a_{i}, \dots, a_{j}\}$. Our trigger mechanism is designed to identify when the performance of the current window is worse than that of previous windows.

As more recently received windows provide more reliable information about the current environment, we adjust the importance of previous windows $\mathcal{W}^{i<j}$ with an exponential decay function. As a result, the weighted average accuracy $\mu_{i<j}$ of previous windows $\mathcal{W}^{i<j}$ can be represented as $\mu_{i<j} = \sum_{i=1}^{j-1} \lambda_i a_i$, where $\lambda_i$ is the normalized weight derived from the exponential decay function that gives more importance to newer windows.

Meanwhile, given the inherent limitations of the student model, fluctuations in accuracy are a common occurrence, which can result in unnecessary triggers. To mitigate this issue, we incorporate a reasonable range of accuracy fluctuations for current window, determined by the standard deviation of previous windows $\sigma_{i<j}$. In summary, we trigger retraining if the following equation is satisfied:
\begin{equation}
  \mu_{i<j} - \sigma_{i<j} > a_j.
  \label{trig}
\end{equation}

Our trigger mechanism ensures that retraining is adaptive to various scenarios and is only initiated when necessary. In cases where the student model performs well and does not require retraining, we store the selected frames and groundtruth labels of this window in the Buffer until next trigger. This allows us to use all the data from the received windows for next retraining. Additionally, the Buffer is cleared after each retraining cycle.

\subsection{Retraining Manager}\label{man}
\begin{figure}[t]
    \centering
    \begin{minipage}[t]{0.22\textwidth}
      \centering
      \includegraphics[width=\textwidth]{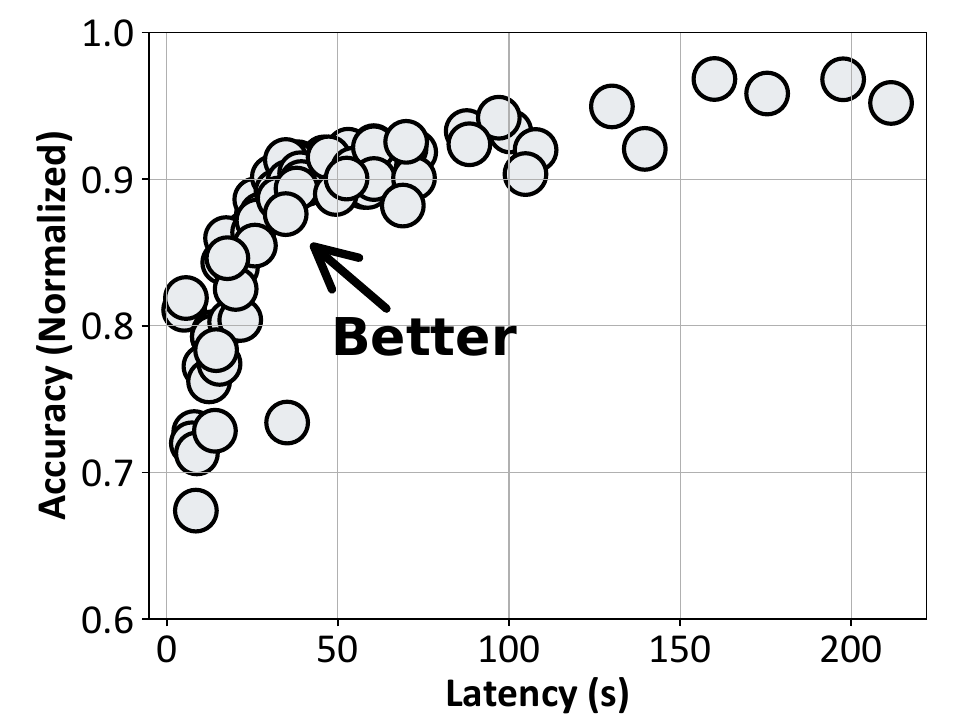}
      \caption{Accuracy and latency performance of different configurations.}
      \label{allconfig}
    \end{minipage}
    \hfill
    \begin{minipage}[t]{0.22\textwidth}
      \centering
      \includegraphics[width=\textwidth]{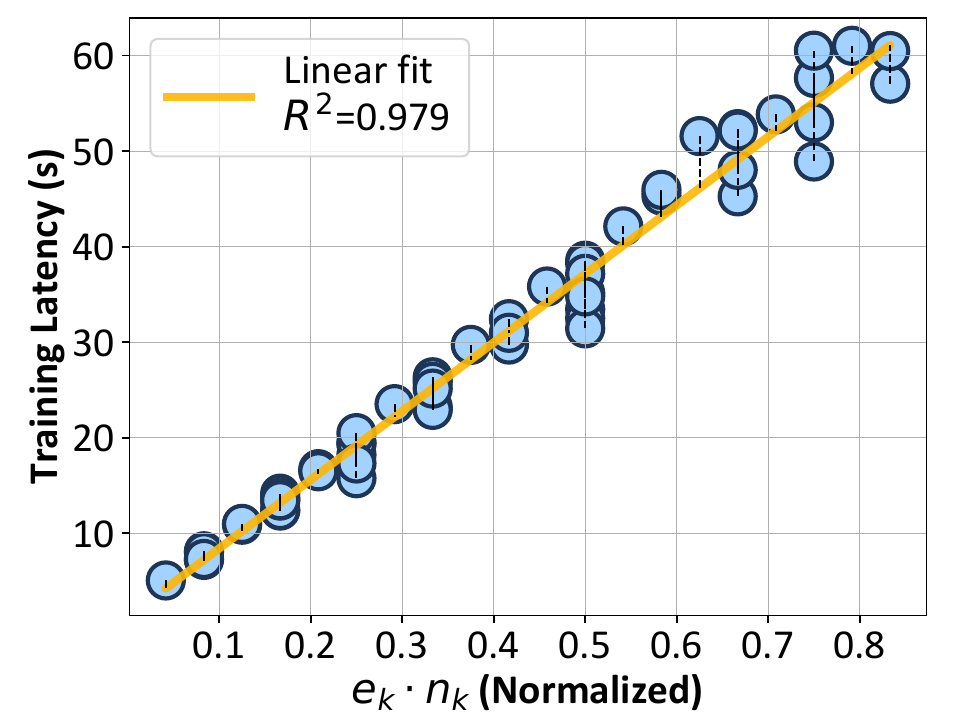}
      \caption{$e_k \cdot n_k$ is approximately linear to the retraining processing latency.}
      \label{linear}
    \end{minipage}
\vspace{-4mm}  
\end{figure}

Upon receiving the signal from Trigger Controller, the Retraining Manager identifies the proper configurations to retrain with the data from both the current window and the Buffer, aiming to achieve high accuracy of the retrained model while minimizing the total retraining time cost to prevent the model from being obsolete. Although accuracy-optimal configurations can enhance performance, they also result in a higher time cost, as illustrated in Fig. \ref{allconfig}. Additionally, we observe that there is only a marginal improvement in accuracy when the time cost is significantly higher, which necessitates the selection of the best configurations. 

We specifically focus on three retraining knobs that play critical roles in both accuracy and time cost: epoch number $e$, frame amount $n$, and teacher model used for knowledge distillation $m$. The $k$-th retraining configurations can be represented as a collection of these factors $\mathbf{c_k} = (e_k, n_k, m_k)$. Then we devise a utility function for each retraining task to evaluate the trade-off of different configurations between accuracy and time overhead, which is defined as: 
\begin{equation}
	U_{k}=A_{k} - \eta_{k} T_{k},\quad \forall k\in K.
	\label{utility}
\end{equation}
where $A_{k}$ represents the accuracy of $k$-th retrained model and is related to different configurations of $\mathbf{c_k}$. $T_{k}$ denotes the total time cost for $k$-th retraining and $\eta_{k}$ profiles the current urgency degree.

\textbf{Latency Model.} The total time overhead $T_k$ for $k$-th retraining can be divided into four parts: \emph{Part. (1)} data transmission latency, \emph{Part. (2)} groundtruth label generation latency, \emph{Part. (3)} training processing latency and \emph{Part. (4)} model transmission latency. We ignore the time cost for Retraining Manager to generate results as we will demonstrate in Section \ref{Evaluation} that this cost can be neglected with our proposed algorithm. Besides, as we have stored data of previous windows in the Buffer, we ignore \emph{Part. (1)} and \emph{Part. (2)} latency from these windows and only consider these two parts latency caused by the trigger window $W_{k}^{\Gamma }$. Hence, the total latency consisting of four parts is modeled as:
\begin{equation}
	T_{k}=\frac{d_{0} |\mathcal S_{k}^{\Gamma }| }{B_{up}} + t_{0} |\mathcal S_{k}^{\Gamma }| + e_{k} n_{k} t(m_{k}) + \frac{M_{0}}{B_{down}},\quad \forall k\in  K.
	\label{timecost}
\end{equation}
where $d_{0}$ and $t_{0}$ are two constants representing the data size and the groundtruth label generation time for each frame, respectively. $M_{0}$ is the size of the student model. $|\mathcal S_{k}^{\Gamma }|$ denotes the number of frames in the trigger window $W_{k}^{\Gamma }$. $B_{up}, B_{down}$ represent the available uplink and downlink bandwidth during transmission, respectively. The \emph{Part. (3)} training processing latency is modeled as $e_{k} n_{k} t(m_{k})$, since we find a significant linear correlation between $e_{k} \cdot n_{k}$ and the training processing latency, as shown in Fig. \ref{linear}. Here, $t(m_{k})$ represents the slope value which varies with different teacher models.

\textbf{Urgency Degree.} The accuracy of student model can be affected differently by various environmental conditions as minor changes can result in a slight decrease, while severe changes can lead to a significant drop, which calls for a faster retraining. As in Step 1 of Algorithm 1, we have obtained the low confidence frame set $\mathcal {\widetilde{\mathcal S}}$ for each window, a larger number of frames in this set indicates a more challenging environment for the student model. Therefore, we utilize the ratio of low confidence frames in the trigger window $W_k^{\Gamma}$ to represent the urgency degree ($\eta_{k}$) for $k$-th retraining. A higher value of $\eta_{k}$ indicates a greater urgency to retrain the current student model. The urgency degree is defined as:

\begin{equation}
	\eta_{k} = \frac{| {\widetilde{\mathcal {S}_k^{\Gamma}}}|}{| \mathcal F|} , \quad \forall k\in  K.
	\label{urgency}
\end{equation}
This definition of urgency degree can help us to make a better tradeoff between accuracy and latency for each retraining.

\textbf{Problem Formulation.} Our objective is to identify the configuration combination of epoch number, frame amount, and teacher model so as to maximize the utility function for all the retraining processes. This problem can be formally formulated as:
\begin{align}
	\mathcal P: \max_{\{\mathbf{e},\mathbf{n},\mathbf{m}\}} \quad & \sum_{k=0}^{ K}U_{k} \\ \label{c1}
	\text{s.t.} \quad & T_{k} \le \tau_{0}, \quad \forall k\in  K \\ \label{c2}
	& e_{k} \le e_{max}, n_{k} \le |\mathcal S_{k}| , \quad \forall k\in  K \\ \label{c3}
    & m_{k} \in \mathcal M, \quad \forall k\in  K 
\end{align}
where Constraint \eqref{c1} limits the total time cost to a threshold $\tau_{0}$, ensuring that the retrained model does not become outdated. Besides, as specified in Constraint \eqref{c2}, the current retraining process must abide by the constraints on the maximum optional frame amount $|\mathcal S_{k}|$ and maximum epoch number $e_{max}$. Constraint \eqref{c3} ensures the selection of the teacher model from the given model set $\mathcal M$. 

Due to this problem is a mixed-integer linear programming (MILP) which is known to be NP-hard, a brute-force approach to solving the problem has a computational complexity of $\mathcal O ( K \cdot |\mathcal M| \cdot | \mathcal S_{k}| \cdot e_{max} )$, making it infeasible to be implemented in practice. Thus, an efficient algorithm should be developed.

\begin{algorithm}[t]
	\caption{Configuration Selection Algorithm}
	\KwIn{ 
        $|\mathcal S_k^{\Gamma} |, |\mathcal S_k |, \eta_k$
            }
	\KwOut{
		$\mathbf{c^*} = [e^*, n^*, m^*]$ : output solution.
            }
        \BlankLine
        {
        $i \gets 0, j \gets 0$ \quad$\triangleright$ initialize iteration counter and fail counter

        $\mathbf{c^*} \gets \mathbf{c_0}$ \quad$\triangleright$ initialize output solution

        $U(\mathbf{c^*}) \gets $getUtility$(\mathbf{c^*};\eta_k;|\mathcal S_k^{\Gamma} |)$ \quad$\triangleright$ compute initial utility score based on Eq.\eqref{utility}\eqref{timecost}
        \BlankLine
        \While{$i < \text{max\_iter}$ and $j < \text{max\_fail}$}{

			$\mathbf{c'}$ $\gets$ generate a new candidate solution randomly within Constraint \eqref{c1}\eqref{c2}\eqref{c3}

            $U(\mathbf{c'}) \gets $getUtility$(\mathbf{c'};\eta_k;|\mathcal S_k^{\Gamma} |)$

            $\Delta f \gets U(\mathbf{c'}) - U(\mathbf{c^*})$ \quad$\triangleright$ compute difference of last solution and new solution

            \If{$\Delta f > 0$ or $\exp(\Delta f / U(\mathbf{c'})) > \text{rand}(0, 1)$}{
                $\mathbf{c^*} \gets \mathbf{c'}$ \quad$\triangleright$ accept new solution

                $j \gets 0$ \quad$\triangleright$ reset fail counter
            }

            \Else{
              $j \gets j + 1$ \quad$\triangleright$ update fail counter 
            }

            
            
            $i \gets i + 1$ \quad$\triangleright$ update iteration counter 
            
        }
        \Return $\mathbf{c^*}$ \quad$\triangleright$ return output solution
        }
\end{algorithm}

\textbf{Algorithm Design.} Instead of considering all retraining tasks simultaneously, we propose Algorithm 2 to identify a sub-optimal configuration for each retraining task individually. At the beginning of the algorithm, we gradually explore the solution space by perturbing the current solution within given constraints. Specifically, each new candidate solution $\mathbf{c'}$ (Line 5) is generated as:

\begin{align}
	e' &= \max(\min(e \pm  randint(0,e_{min}), e_{max}),e_{min}) \\
    n' &= \max(\min(n \pm  randint(0,n_{min}), |\mathcal S_k |),n_{min}) \\
    m' &= randchoice(\mathcal M)
\end{align}

We adopt a new solution $c'$ to update the current optimal solution $c^*$ when its utility score surpasses that of $c^*$. In case the utility score of $c'$ is inferior, we may still adopt it with a probability according to Metropolis Criterion \cite{metro}. We end the search process early if a new candidate solution is not accepted after \emph{max\_fail} times of attempts. Otherwise, we complete all iterations before exiting. Additionally, with the output number of $n^*$, we select the most recent frames as the training set because these frames are better indicators of current environmental conditions.

In summary, through the collaborative efforts among the Key Frame Extractor on end cameras and the Trigger Controller, the Retraining Manager on edge servers, our system can enhance video analytics accuracy under adverse environmental conditions.

\section{Performance Evaluation}\label{Evaluation}

In this section, we demonstrate the efficiency of our system design with real-world implementations and experiments on object detection, a fundamental task in many high-level vision applications.

\begin{figure}[t]
    \centering
    \begin{minipage}[t]{0.22\textwidth}
      \centering
      \includegraphics[width=\textwidth]{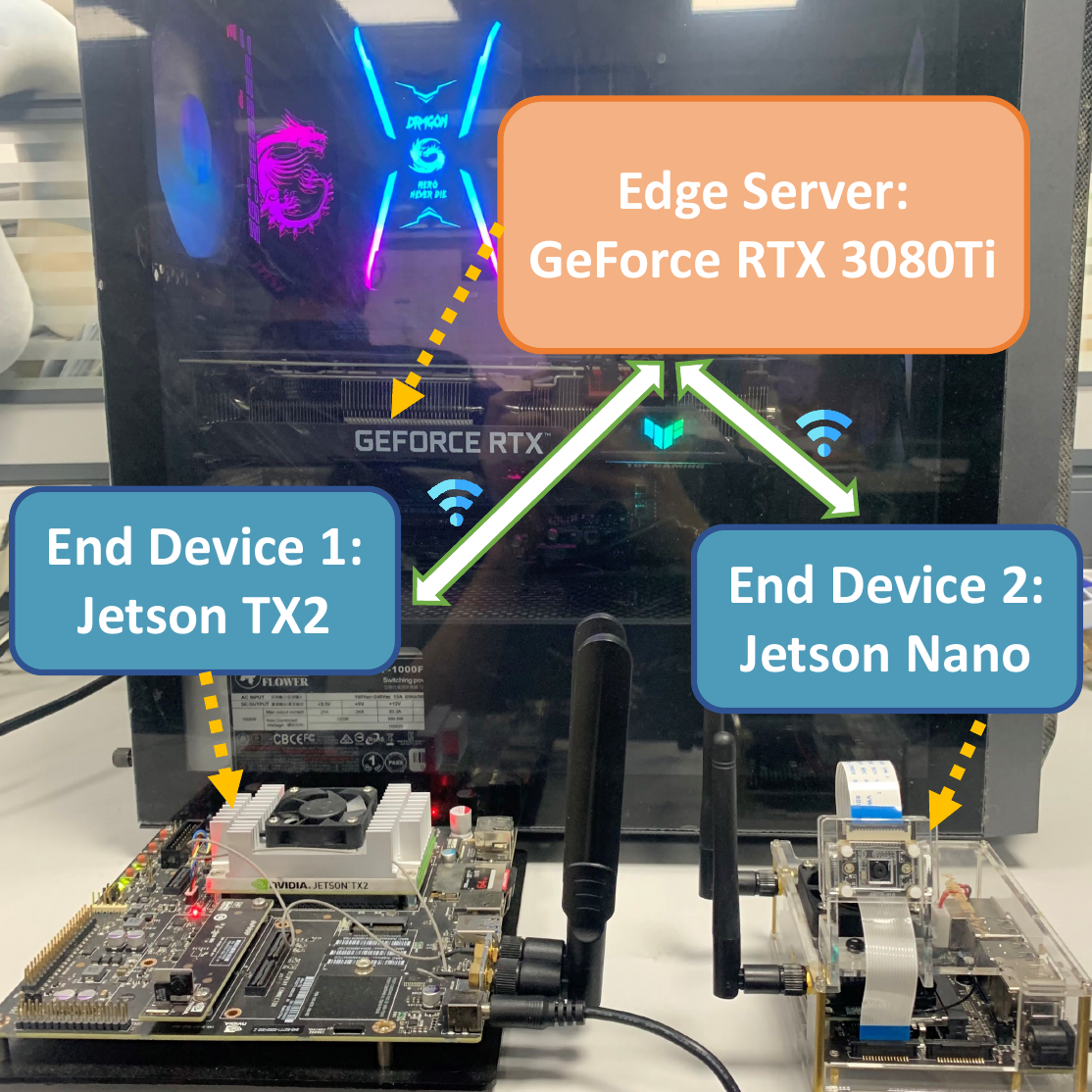}
      \caption{A snapshot of the system prototype implementation.}
      \label{nano}
    \end{minipage}
    \hfill
    \begin{minipage}[t]{0.22\textwidth}
      \centering
      \includegraphics[width=\textwidth]{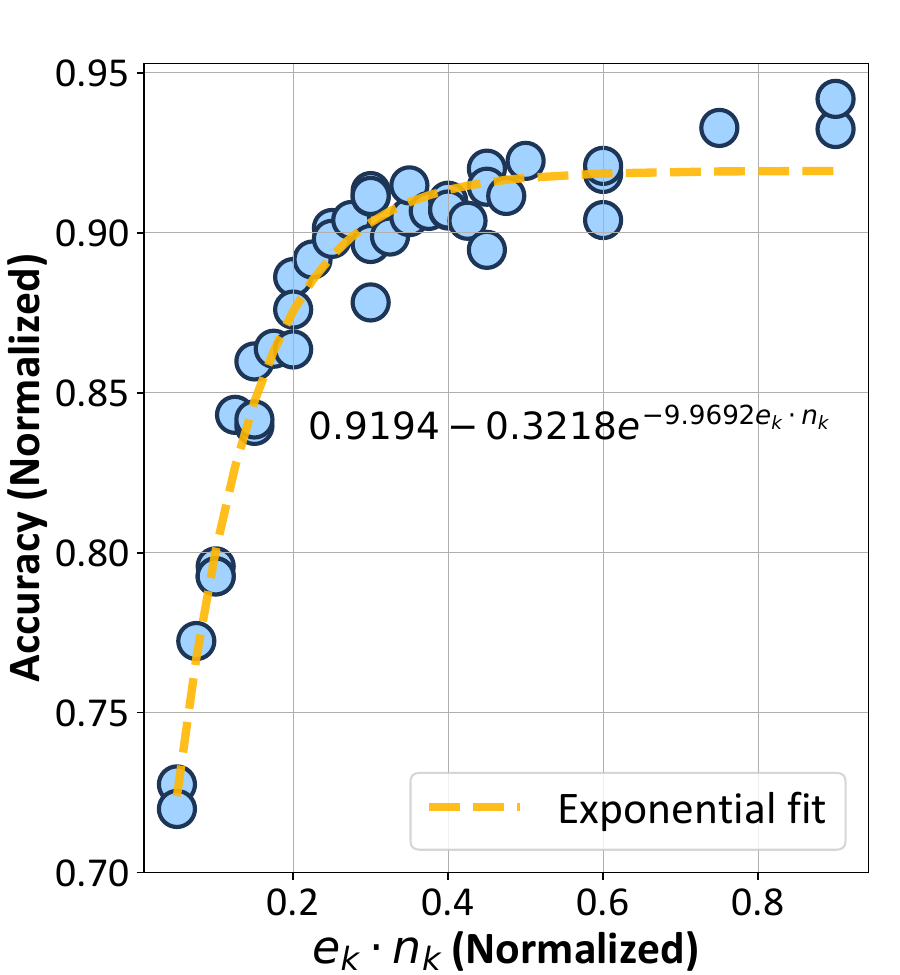}
      \caption{$e_k \cdot n_k$ is fit to accuracy with an exponential function.}
      \label{fit}
    \end{minipage}
  \end{figure}


\subsection{Experimental Settings}

\begin{figure*}[t]
  \centering
  \includegraphics[scale=0.51]{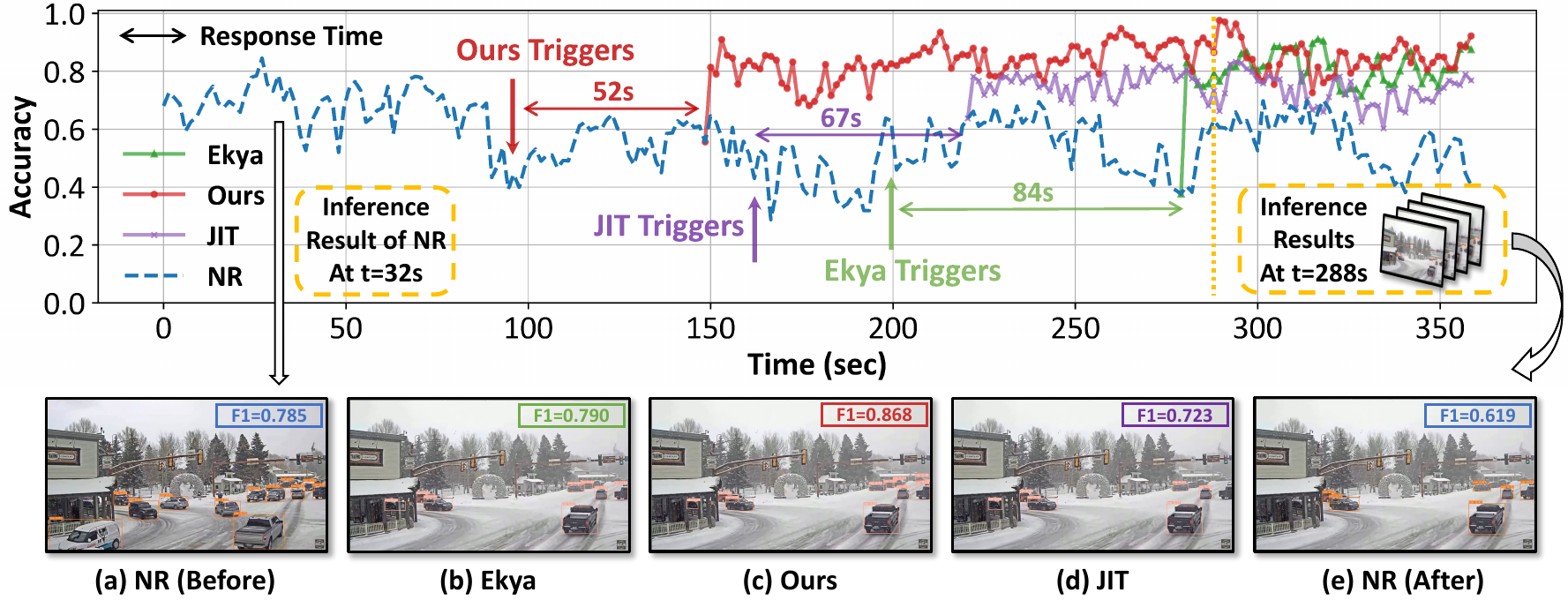}
  \caption{Comparison of the first model update process of different schemes under the Heavy Snow scene.}
  \label{vis}
\end{figure*}

\noindent\textbf{Setup.} We utilize Yolov5 series models \cite{c1} pre-trained on COCO dataset \cite{c4} in our evaluation. We develop a test-bed, consisting of two end devices with different computation resources, \emph{i.e.}, Jetson Nano \cite{nano} and Jetson TX2 \cite{tx2}. Both end devices are connected to the same edge server via wireless links. On the end devices, we implement Yolov5-n as the student model with TensorRT engine \cite{c6} to accelerate inference. A desktop computer equipped with NVIDIA GeForce RTX 3080Ti GPU acts as the edge server, on which the Oracle model Yolov5-X is deployed to conduct retraining tasks. A snapshot of our system prototype is shown in Fig. \ref{nano}.

\noindent\textbf{Datasets.} We evaluate our design under three typical challenging environmental conditions: Low Light, Heavy Snow, and Heavy Rain. These adverse conditions exhibit varying levels of change intensity, with Low Light scene changing most smoothly and Heavy Rain scene changing most dramatically. Due to the limited availability of datasets on various adverse environments, particularly those containing long-time-span videos with ample valid objects, we collect these videos from YouTube\footnote{https://www.youtube.com/watch?v=1EiC9bvVGnk and https://www.youtube.com/watch?v=vSq3eUasKWA}, with a resolution of 1920 $\times$ 1080 and a frame rate of 30 frames per second. 

\noindent\textbf{Parameters.} We set the parameters in our proposed algorithms and components, including $\beta = 0.4$, $W_{length} = 10s$, $\tau_0 = 60s$, $n_{min} = 50$, $e_{min} = 5$ and $e_{max} = 100$. Our teacher model set $\mathcal M$ comprises of three models, namely Yolov5-s, Yolov5-m, and Yolov5-l. By conducting offline profiling, we are able to not only determine the value of $t(m)$ for each teacher model but also establish the relationship between the accuracy and $e_k \cdot n_k$ across three teacher models. An example of the fitted accuracy curve using Yolov5-s as the teacher model is presented in Fig. \ref{fit}.

\noindent\textbf{Schemes.} We compare with the following schemes.
\begin{itemize} 
	\item \textbf{No Retraining (NR).} We run the pre-trained student model directly on end cameras without any specific information from the current video contents.
	\item \textbf{One-Time Retraining (OTR).} We retrain the student model only once with the key frames extracted from the first 60s of the video and test on the rest. This strategy can be seen as our system excluding the component of Trigger Controller. 
	\item \textbf{JIT \cite{JIT}.} JIT employs a fixed accuracy threshold to control retraining and uses a retraining strategy where the student model stops retraining when either the accuracy threshold is reached or the maximum iterations have been executed.
	\item \textbf{Ekya \cite{Ekya}.} Ekya uses a fixed retraining window of 200s, \emph{i.e.}, updates the student model periodically. Ekya selects the best configurations for retraining through an online micro-profiler, which runs a few extra epochs on a small fraction of training data to build a profile of the training curve for each configuration. In Ekya, both inference and retraining are done on the edge server, indicating that the whole video stream will be continuously offloaded from the camera to the edge server for real-time inference purpose. 
\end{itemize}

\subsection{End-to-End Performance}
\textbf{Performance of first model update.} We compare the first model update process of different schemes under Heavy Snow conditions in Fig. \ref{vis}. Before the occurrence of adverse environments, the student model is not retrained but performs well as the inference result of NR at $t=32s$ shows in Fig. \ref{vis}(a). Then the heavy snow happens and we compare the detection results of four schemes on the same frame extracted under bad environments at $t=295s$ in Fig. \ref{vis}(b)-\ref{vis}(e) respectively. As presented in Fig. \ref{vis}(e), the previous student model fails to adapt to the degraded environmental conditions with a large accuracy drop. Instead, our adaptive Trigger Controller makes our system the first to detect environmental deterioration and initiate retraining. Besides, our efficient configuration selection algorithm enables our system to retrain and update the student model in just 52s, resulting in a quicker response time and higher overall accuracy. In contrast, JIT and Ekya trigger retraining later than ours and take 67s and 84s to complete the whole retraining process respectively, resulting in a longer response time during which the accuracy maintains low.

\begin{figure}[t]
  \centering
  \includegraphics[scale=0.45]{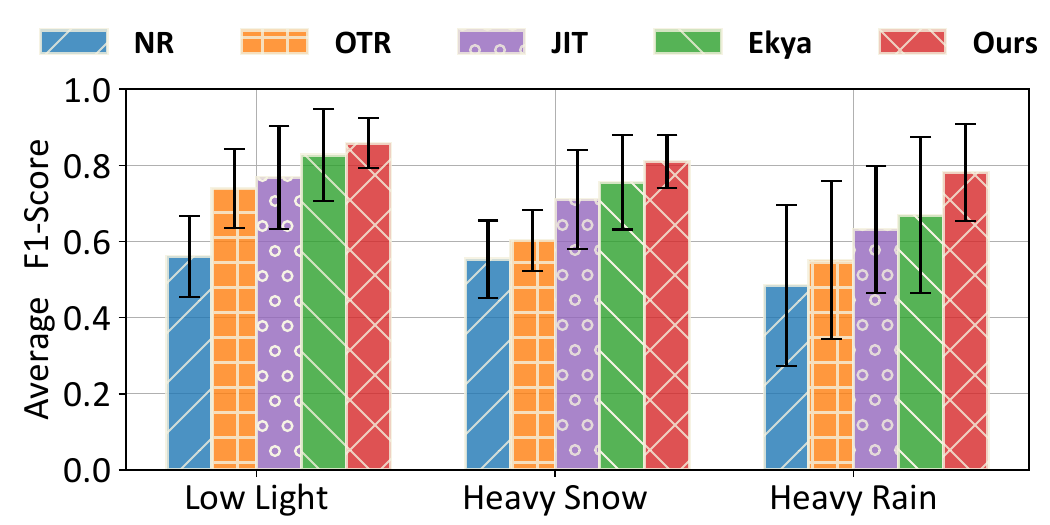}
  \caption{Average F1-Score and standard deviation.}
  \label{sigma}
\end{figure}

\textbf{Overall accuracy.} The average F1-Score results and corresponding standard deviation values under three different scenes are presented in Fig. \ref{sigma}. Our system outperforms all the other methods in terms of accuracy, achieving up to 24\% improvements compared to the naive NR strategy. Besides, our system demonstrates a low incidence of sudden accuracy drop due to our timely trigger of retraining when accuracy falls below the expected level. As a result, our method exhibits the lowest standard deviation values across all scenarios, indicating the robustness of our system. 

OTR can significantly improve accuracy under Low Light scene, which has a relatively smooth and gradual change in environmental conditions. However, in more volatile environments like Heavy Rain, this approach is less effective because the student model cannot adapt to new changes in environments. JIT triggers retraining when the accuracy threshold is reached, but this fixed threshold is not optimized for different environmental conditions, preventing JIT from gaining better accuracy in more challenging scenes. 

Ekya is able to select appropriate configurations for retraining with the help of an online micro-profiler, which works well under less challenging scenes like Low Light. Nevertheless, when confronted with the worse Heavy Rain scene, the micro-profiler fails to profile the training curve correctly due to the difficulty of achieving convergence within just a few epochs, which ultimately leads to inferior output configurations. Besides, the fixed update interval also prevents Ekya from responding to environmental deterioration timely. Compared to Ekya, our method achieves 12\% higher accuracy and maintains good performance in all challenging scenes.

\begin{figure}[t]
    \centering
    \includegraphics[scale=0.27]{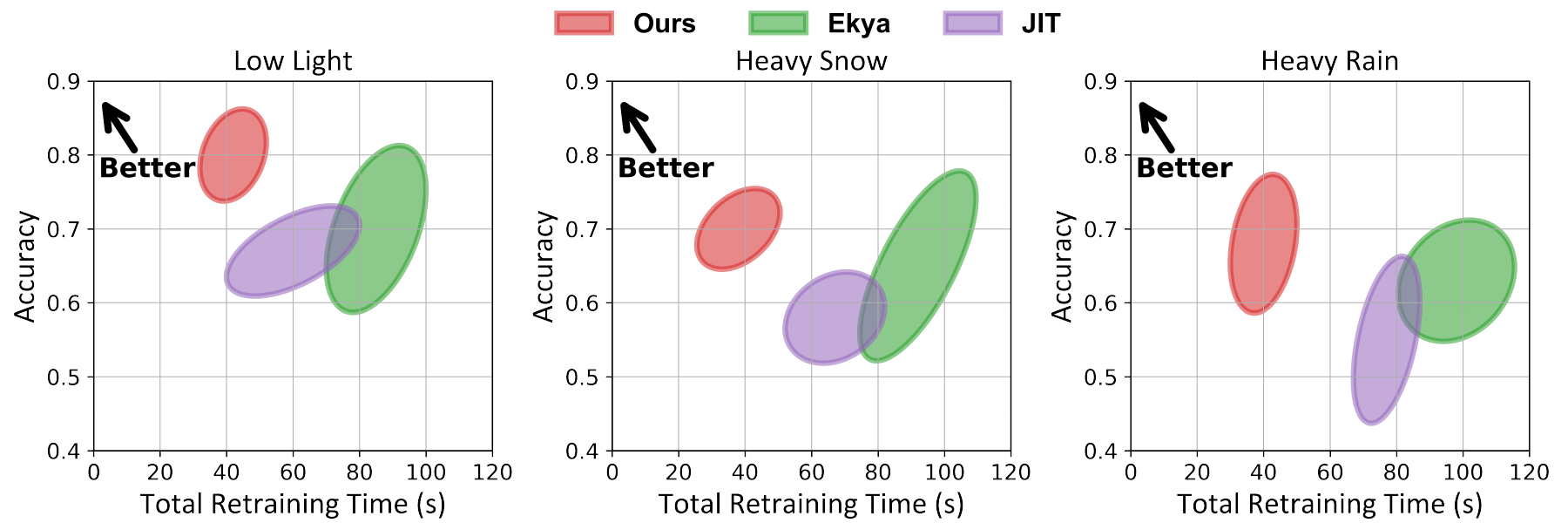}
    \caption{Accuracy v.s. the total retraining time overhead.}
    \label{circle}
\end{figure}

\textbf{Accuracy-latency tradeoff.} By jointly considering accuracy and total retraining time cost, we demonstrate our system is capable of striking a balance between both metrics, as shown in Fig. \ref{circle}. We achieve better accuracy and latency tradeoff than Ekya and JIT across all three scenes due to our efficient configuration selection algorithm, which completes in just 20ms. In contrast, JIT and Ekya take more time to retrain but do not achieve better accuracy due to the deficiency of their retraining strategies. For example, JIT aggressively retrains the model to the pre-defined accuracy level, while Ekya requires additional micro-training to obtain a training curve for configuration selection, both of which have an inferior balance of accuracy and time cost. Besides, longer retraining time also increases the risk of the retrained model becoming outdated, leading to poor performance as the video content changes.

\textbf{Impact of bandwidth condition.} We illustrate our system is robust to bandwidth variations in Fig. \ref{bw} under the most challenging Heavy Rain scene. Despite unfavorable bandwidth conditions, our system only experiences a 5\% decrease in accuracy. This is because our Key Frame Extractor can select the most representative frames for retraining within current bandwidth constraints. NR and OTR are not sensitive to bandwidth conditions because of none or few communication with the edge. JIT transmits recent frames to the edge with different strides, but it does not consider the transmission under limited bandwidth conditions, leading to accuracy degradation. As Ekya is designed to perform both inference and retraining tasks on the edge server without any video-to-edge offloading specification, we assume it adopts the typical naive offloading strategy, which continuously transmits a small video slot from camera to edge. The result shows that this strategy is most vulnerable to bandwidth variations, with up to 10\% accuracy degradation.

\subsection{Deep Dive of Retraining Performance}

\begin{figure}[t]
  \centering
  \begin{minipage}[t]{0.23\textwidth}
    \centering
    \includegraphics[width=\textwidth]{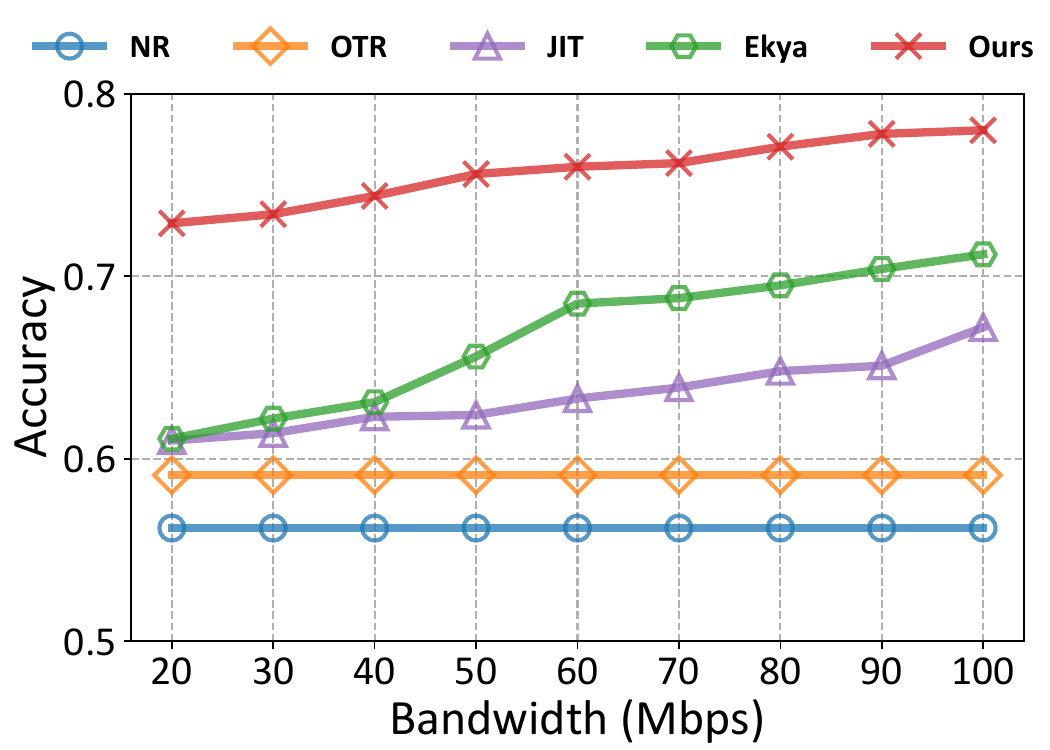}
    \caption{Impact of bandwidth condition.}
    \label{bw}
  \end{minipage}
  \hfill
  \begin{minipage}[t]{0.23\textwidth}
    \centering
    \includegraphics[width=\textwidth]{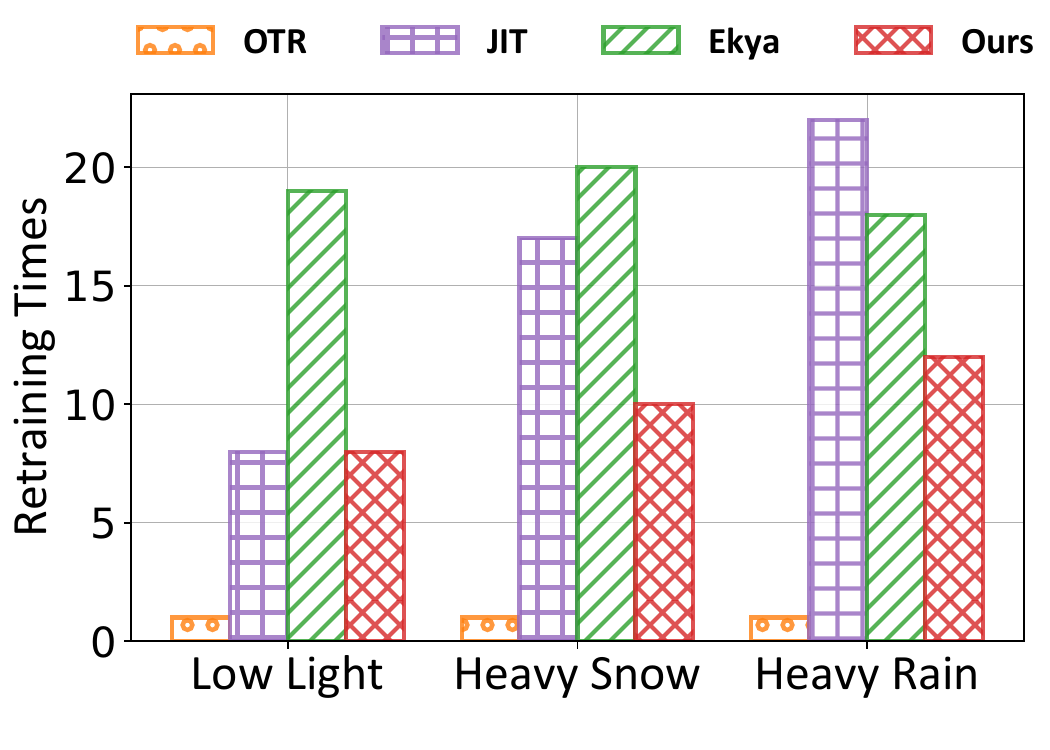}
    \caption{Comparison of retraining times.}
    \label{times}
  \end{minipage}
\end{figure}

\textbf{Retraining times.} As shown in Fig. \ref{times}, our method can avoid unnecessary retraining effectively across all three scenes. JIT employs a fixed accuracy threshold for model updates, which can cause frequent triggers in challenging scenes like Heavy Rain, where accuracy tends to suffer. Ekya adopts a fixed retraining window, which leads to nearly twice the number of retraining times compared to our approach under Low Light Scene. The reduction in retraining times is highly meaningful, as fewer retraining can significantly save the computational resources on the edge server.

\begin{figure}[t]
  \centering
  \includegraphics[scale=0.4]{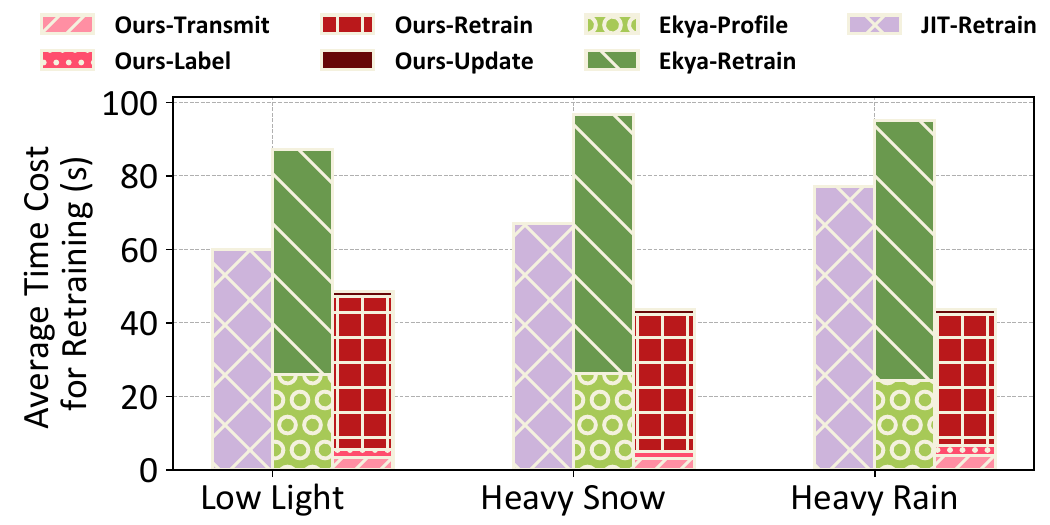}
  \caption{Decomposition of total retraining time overhead.}
  \label{avgretrain}
\end{figure}

\textbf{Decomposition of total retraining time cost.} As mentioned in Section \ref{man}, our total latency for each retraining process consists of four parts: transmission, labeling, training, and updating. Our proposed algorithm enables us to select configurations achieving both high accuracy and low time cost. A lower retraining time cost can prevent the retrained model from becoming obsolete and save computing resources in the meantime. JIT aggressively retrains the model until the expected accuracy or max iterations, resulting in an unstable time cost. Ekya adopts an online micro-profiler, which brings extra time costs. Fig. \ref{avgretrain} compares different parts of the total retraining time cost. We cut the retraining time by 52.7\% compared to Ekya while still achieving superior performance. Besides, our time cost for the four parts accounts for 7.8\%, 5.2\%, 86.2\%, and 0.8\% respectively, with the three non-training parts (\emph{i.e.}, transmission, labeling, and updating) occupying a relatively small fraction. However, Ekya's micro-profiler takes up 27\% of the total runtime (\emph{i.e.}, 25s), hindering it from promptly adapting to new variations in environmental conditions.

\section{Conclusion}\label{Conclusion}
In this paper, we have proposed an edge-assisted approach to continuously update lightweight models on resource-constrained end cameras, aiming to improve the performance of real-time video analytics under adverse environmental conditions. We have devised a Key Frame Extractor to select the most informative frames under bandwidth constraints, an adaptive Trigger Controller to inform model updates promptly, and a Retraining Manager to select configurations for efficient retraining. We have implemented a test-bed and experimental results demonstrated that our system achieves better accuracy under all tested adverse environments with time cost much less than current benchmarks. For the future work, we will focus on further reducing bandwidth consumption by incorporating the encoding process.


\bibliographystyle{ACM-Reference-Format}
\balance
\bibliography{mmref}


\begin{thebibliography}{58}


\ifx \showCODEN    \undefined \def \showCODEN     #1{\unskip}     \fi
\ifx \showDOI      \undefined \def \showDOI       #1{#1}\fi
\ifx \showISBNx    \undefined \def \showISBNx     #1{\unskip}     \fi
\ifx \showISBNxiii \undefined \def \showISBNxiii  #1{\unskip}     \fi
\ifx \showISSN     \undefined \def \showISSN      #1{\unskip}     \fi
\ifx \showLCCN     \undefined \def \showLCCN      #1{\unskip}     \fi
\ifx \shownote     \undefined \def \shownote      #1{#1}          \fi
\ifx \showarticletitle \undefined \def \showarticletitle #1{#1}   \fi
\ifx \showURL      \undefined \def \showURL       {\relax}        \fi
\providecommand\bibfield[2]{#2}
\providecommand\bibinfo[2]{#2}
\providecommand\natexlab[1]{#1}
\providecommand\showeprint[2][]{arXiv:#2}

\bibitem[Ananthanarayanan et~al\mbox{.}(2017)]%
        {a1}
\bibfield{author}{\bibinfo{person}{Ganesh Ananthanarayanan},
  \bibinfo{person}{Paramvir Bahl}, \bibinfo{person}{Peter Bod{\'\i}k},
  \bibinfo{person}{Krishna Chintalapudi}, \bibinfo{person}{Matthai Philipose},
  \bibinfo{person}{Lenin Ravindranath}, {and} \bibinfo{person}{Sudipta Sinha}.}
  \bibinfo{year}{2017}\natexlab{}.
\newblock \showarticletitle{Real-time video analytics: The killer app for edge
  computing}.
\newblock \bibinfo{journal}{\emph{computer}} \bibinfo{volume}{50},
  \bibinfo{number}{10} (\bibinfo{year}{2017}), \bibinfo{pages}{58--67}.
\newblock


\bibitem[Bhardwaj et~al\mbox{.}(2022)]%
        {Ekya}
\bibfield{author}{\bibinfo{person}{Romil Bhardwaj}, \bibinfo{person}{Zhengxu
  Xia}, \bibinfo{person}{Ganesh Ananthanarayanan}, \bibinfo{person}{Junchen
  Jiang}, \bibinfo{person}{Yuanchao Shu}, \bibinfo{person}{Nikolaos
  Karianakis}, \bibinfo{person}{Kevin Hsieh}, \bibinfo{person}{Paramvir Bahl},
  {and} \bibinfo{person}{Ion Stoica}.} \bibinfo{year}{2022}\natexlab{}.
\newblock \showarticletitle{Ekya: Continuous learning of video analytics models
  on edge compute servers}. In \bibinfo{booktitle}{\emph{19th USENIX Symposium
  on Networked Systems Design and Implementation (NSDI 22)}}.
  \bibinfo{pages}{119--135}.
\newblock


\bibitem[Canel et~al\mbox{.}(2019)]%
        {scale2}
\bibfield{author}{\bibinfo{person}{Christopher Canel}, \bibinfo{person}{Thomas
  Kim}, \bibinfo{person}{Giulio Zhou}, \bibinfo{person}{Conglong Li},
  \bibinfo{person}{Hyeontaek Lim}, \bibinfo{person}{David~G Andersen},
  \bibinfo{person}{Michael Kaminsky}, {and} \bibinfo{person}{Subramanya
  Dulloor}.} \bibinfo{year}{2019}\natexlab{}.
\newblock \showarticletitle{Scaling video analytics on constrained edge nodes}.
\newblock \bibinfo{journal}{\emph{Proceedings of Machine Learning and Systems}}
   \bibinfo{volume}{1} (\bibinfo{year}{2019}), \bibinfo{pages}{406--417}.
\newblock


\bibitem[Chen and Ran(2019)]%
        {review}
\bibfield{author}{\bibinfo{person}{Jiasi Chen} {and} \bibinfo{person}{Xukan
  Ran}.} \bibinfo{year}{2019}\natexlab{}.
\newblock \showarticletitle{Deep learning with edge computing: A review}.
\newblock \bibinfo{journal}{\emph{Proc. IEEE}} \bibinfo{volume}{107},
  \bibinfo{number}{8} (\bibinfo{year}{2019}), \bibinfo{pages}{1655--1674}.
\newblock


\bibitem[Chen et~al\mbox{.}(2015)]%
        {glimpse}
\bibfield{author}{\bibinfo{person}{Tiffany Yu-Han Chen}, \bibinfo{person}{Lenin
  Ravindranath}, \bibinfo{person}{Shuo Deng}, \bibinfo{person}{Paramvir Bahl},
  {and} \bibinfo{person}{Hari Balakrishnan}.} \bibinfo{year}{2015}\natexlab{}.
\newblock \showarticletitle{Glimpse: Continuous, real-time object recognition
  on mobile devices}. In \bibinfo{booktitle}{\emph{Proceedings of the 13th ACM
  Conference on Embedded Networked Sensor Systems}}. \bibinfo{pages}{155--168}.
\newblock


\bibitem[Chen et~al\mbox{.}(2021)]%
        {flat}
\bibfield{author}{\bibinfo{person}{Xu Chen}, \bibinfo{person}{Chenqiang Gao},
  \bibinfo{person}{Feng Yang}, \bibinfo{person}{Xiaohan Wang},
  \bibinfo{person}{Yi Yang}, {and} \bibinfo{person}{Yahong Han}.}
  \bibinfo{year}{2021}\natexlab{}.
\newblock \showarticletitle{Video-to-Image casting: A flatting method for video
  analysis}. In \bibinfo{booktitle}{\emph{Proceedings of the 29th ACM
  International Conference on Multimedia}}. \bibinfo{pages}{4958--4966}.
\newblock


\bibitem[Cheng et~al\mbox{.}(2023)]%
        {yan}
\bibfield{author}{\bibinfo{person}{Yan Cheng}, \bibinfo{person}{Peng Yang},
  \bibinfo{person}{Ning Zhang}, {and} \bibinfo{person}{Jiawei Hou}.}
  \bibinfo{year}{2023}\natexlab{}.
\newblock \showarticletitle{Edge-Assisted Lightweight Region-of-Interest
  Extraction and Transmission for Vehicle Perception}. In
  \bibinfo{booktitle}{\emph{GLOBECOM 2023-2023 IEEE Global Communications
  Conference}}. IEEE, \bibinfo{pages}{to appear}.
\newblock


\bibitem[Dai et~al\mbox{.}(2022)]%
        {a3}
\bibfield{author}{\bibinfo{person}{Xiangxiang Dai}, \bibinfo{person}{Peng
  Yang}, \bibinfo{person}{Xinyu Zhang}, \bibinfo{person}{Zhewei Dai}, {and}
  \bibinfo{person}{Li Yu}.} \bibinfo{year}{2022}\natexlab{}.
\newblock \showarticletitle{RESPIRE: Reducing Spatial--Temporal Redundancy for
  Efficient Edge-Based Industrial Video Analytics}.
\newblock \bibinfo{journal}{\emph{IEEE Transactions on Industrial Informatics}}
  \bibinfo{volume}{18}, \bibinfo{number}{12} (\bibinfo{year}{2022}),
  \bibinfo{pages}{9324--9334}.
\newblock


\bibitem[Du et~al\mbox{.}(2020)]%
        {dds}
\bibfield{author}{\bibinfo{person}{Kuntai Du}, \bibinfo{person}{Ahsan Pervaiz},
  \bibinfo{person}{Xin Yuan}, \bibinfo{person}{Aakanksha Chowdhery},
  \bibinfo{person}{Qizheng Zhang}, \bibinfo{person}{Henry Hoffmann}, {and}
  \bibinfo{person}{Junchen Jiang}.} \bibinfo{year}{2020}\natexlab{}.
\newblock \showarticletitle{Server-driven video streaming for deep learning
  inference}. In \bibinfo{booktitle}{\emph{Proceedings of the Annual conference
  of the ACM Special Interest Group on Data Communication on the applications,
  technologies, architectures, and protocols for computer communication}}.
  \bibinfo{pages}{557--570}.
\newblock


\bibitem[Guo et~al\mbox{.}(2021)]%
        {crossroi}
\bibfield{author}{\bibinfo{person}{Hongpeng Guo}, \bibinfo{person}{Shuochao
  Yao}, \bibinfo{person}{Zhe Yang}, \bibinfo{person}{Qian Zhou}, {and}
  \bibinfo{person}{Klara Nahrstedt}.} \bibinfo{year}{2021}\natexlab{}.
\newblock \showarticletitle{CrossRoI: Cross-camera region of interest
  optimization for efficient real time video analytics at scale}. In
  \bibinfo{booktitle}{\emph{Proceedings of the 12th ACM Multimedia Systems
  Conference}}. \bibinfo{pages}{186--199}.
\newblock


\bibitem[Han et~al\mbox{.}(2020)]%
        {b1}
\bibfield{author}{\bibinfo{person}{Kai Han}, \bibinfo{person}{Yunhe Wang},
  \bibinfo{person}{Qi Tian}, \bibinfo{person}{Jianyuan Guo},
  \bibinfo{person}{Chunjing Xu}, {and} \bibinfo{person}{Chang Xu}.}
  \bibinfo{year}{2020}\natexlab{}.
\newblock \showarticletitle{Ghostnet: More features from cheap operations}. In
  \bibinfo{booktitle}{\emph{Proceedings of the IEEE/CVF conference on computer
  vision and pattern recognition}}. \bibinfo{pages}{1580--1589}.
\newblock


\bibitem[He et~al\mbox{.}(2021)]%
        {vod}
\bibfield{author}{\bibinfo{person}{Lu He}, \bibinfo{person}{Qianyu Zhou},
  \bibinfo{person}{Xiangtai Li}, \bibinfo{person}{Li Niu},
  \bibinfo{person}{Guangliang Cheng}, \bibinfo{person}{Xiao Li},
  \bibinfo{person}{Wenxuan Liu}, \bibinfo{person}{Yunhai Tong},
  \bibinfo{person}{Lizhuang Ma}, {and} \bibinfo{person}{Liqing Zhang}.}
  \bibinfo{year}{2021}\natexlab{}.
\newblock \showarticletitle{End-to-end video object detection with
  spatial-temporal transformers}. In \bibinfo{booktitle}{\emph{Proceedings of
  the 29th ACM International Conference on Multimedia}}.
  \bibinfo{pages}{1507--1516}.
\newblock


\bibitem[He et~al\mbox{.}(2023)]%
        {yuan}
\bibfield{author}{\bibinfo{person}{Yuanyi He}, \bibinfo{person}{Peng Yang},
  \bibinfo{person}{Tian Qin}, {and} \bibinfo{person}{Ning Zhang}.}
  \bibinfo{year}{2023}\natexlab{}.
\newblock \showarticletitle{End-Edge Coordinated Joint Encoding and Neural
  Enhancement for Low-Light Video Analytics}. In
  \bibinfo{booktitle}{\emph{GLOBECOM 2023-2023 IEEE Global Communications
  Conference}}. \bibinfo{pages}{to appear}.
\newblock


\bibitem[Hou et~al\mbox{.}(2023)]%
        {jerry}
\bibfield{author}{\bibinfo{person}{Jiawei Hou}, \bibinfo{person}{Peng Yang},
  \bibinfo{person}{Tian Qin}, {and} \bibinfo{person}{Wen Wu}.}
  \bibinfo{year}{2023}\natexlab{}.
\newblock \showarticletitle{Edge-Coordinated On-Road Perception for Connected
  Autonomous Vehicles Using Point Cloud}. In
  \bibinfo{booktitle}{\emph{Proceedings of 31st Biennial Symposium on
  Communications}}. \bibinfo{pages}{to appear}.
\newblock


\bibitem[Hung et~al\mbox{.}(2018)]%
        {videoedge}
\bibfield{author}{\bibinfo{person}{Chien-Chun Hung}, \bibinfo{person}{Ganesh
  Ananthanarayanan}, \bibinfo{person}{Peter Bodik}, \bibinfo{person}{Leana
  Golubchik}, \bibinfo{person}{Minlan Yu}, \bibinfo{person}{Paramvir Bahl},
  {and} \bibinfo{person}{Matthai Philipose}.} \bibinfo{year}{2018}\natexlab{}.
\newblock \showarticletitle{Videoedge: Processing camera streams using
  hierarchical clusters}. In \bibinfo{booktitle}{\emph{2018 IEEE/ACM Symposium
  on Edge Computing (SEC)}}. IEEE, \bibinfo{pages}{115--131}.
\newblock


\bibitem[Jain et~al\mbox{.}(2019)]%
        {scale}
\bibfield{author}{\bibinfo{person}{Samvit Jain}, \bibinfo{person}{Ganesh
  Ananthanarayanan}, \bibinfo{person}{Junchen Jiang}, \bibinfo{person}{Yuanchao
  Shu}, {and} \bibinfo{person}{Joseph Gonzalez}.}
  \bibinfo{year}{2019}\natexlab{}.
\newblock \showarticletitle{Scaling video analytics systems to large camera
  deployments}. In \bibinfo{booktitle}{\emph{Proceedings of the 20th
  International Workshop on Mobile Computing Systems and Applications}}.
  \bibinfo{pages}{9--14}.
\newblock


\bibitem[Jain et~al\mbox{.}(2020)]%
        {spatula}
\bibfield{author}{\bibinfo{person}{Samvit Jain}, \bibinfo{person}{Xun Zhang},
  \bibinfo{person}{Yuhao Zhou}, \bibinfo{person}{Ganesh Ananthanarayanan},
  \bibinfo{person}{Junchen Jiang}, \bibinfo{person}{Yuanchao Shu},
  \bibinfo{person}{Paramvir Bahl}, {and} \bibinfo{person}{Joseph Gonzalez}.}
  \bibinfo{year}{2020}\natexlab{}.
\newblock \showarticletitle{Spatula: Efficient cross-camera video analytics on
  large camera networks}. In \bibinfo{booktitle}{\emph{2020 IEEE/ACM Symposium
  on Edge Computing (SEC)}}. IEEE, \bibinfo{pages}{110--124}.
\newblock


\bibitem[Jiang et~al\mbox{.}(2018)]%
        {chame}
\bibfield{author}{\bibinfo{person}{Junchen Jiang}, \bibinfo{person}{Ganesh
  Ananthanarayanan}, \bibinfo{person}{Peter Bodik}, \bibinfo{person}{Siddhartha
  Sen}, {and} \bibinfo{person}{Ion Stoica}.} \bibinfo{year}{2018}\natexlab{}.
\newblock \showarticletitle{Chameleon: scalable adaptation of video analytics}.
  In \bibinfo{booktitle}{\emph{Proceedings of the 2018 conference of the ACM
  special interest group on data communication}}. \bibinfo{pages}{253--266}.
\newblock


\bibitem[Kang et~al\mbox{.}(2017)]%
        {noscope}
\bibfield{author}{\bibinfo{person}{Daniel Kang}, \bibinfo{person}{John Emmons},
  \bibinfo{person}{Firas Abuzaid}, \bibinfo{person}{Peter Bailis}, {and}
  \bibinfo{person}{Matei Zaharia}.} \bibinfo{year}{2017}\natexlab{}.
\newblock \showarticletitle{Noscope: optimizing neural network queries over
  video at scale}.
\newblock \bibinfo{journal}{\emph{arXiv preprint arXiv:1703.02529}}
  (\bibinfo{year}{2017}).
\newblock


\bibitem[Khani et~al\mbox{.}(2023)]%
        {recl}
\bibfield{author}{\bibinfo{person}{Mehrdad Khani}, \bibinfo{person}{Ganesh
  Ananthanarayanan}, \bibinfo{person}{Kevin Hsieh}, \bibinfo{person}{Junchen
  Jiang}, \bibinfo{person}{Ravi Netravali}, \bibinfo{person}{Yuanchao Shu},
  \bibinfo{person}{Mohammad Alizadeh}, {and} \bibinfo{person}{Victor Bahl}.}
  \bibinfo{year}{2023}\natexlab{}.
\newblock \showarticletitle{$\{$RECL$\}$: Responsive $\{$Resource-Efficient$\}$
  Continuous Learning for Video Analytics}. In \bibinfo{booktitle}{\emph{20th
  USENIX Symposium on Networked Systems Design and Implementation (NSDI 23)}}.
  \bibinfo{pages}{917--932}.
\newblock


\bibitem[Khani et~al\mbox{.}(2021)]%
        {AMS}
\bibfield{author}{\bibinfo{person}{Mehrdad Khani}, \bibinfo{person}{Pouya
  Hamadanian}, \bibinfo{person}{Arash Nasr-Esfahany}, {and}
  \bibinfo{person}{Mohammad Alizadeh}.} \bibinfo{year}{2021}\natexlab{}.
\newblock \showarticletitle{Real-time video inference on edge devices via
  adaptive model streaming}. In \bibinfo{booktitle}{\emph{Proceedings of the
  IEEE/CVF International Conference on Computer Vision}}.
  \bibinfo{pages}{4572--4582}.
\newblock


\bibitem[Li et~al\mbox{.}(2019)]%
        {rilod}
\bibfield{author}{\bibinfo{person}{Dawei Li}, \bibinfo{person}{Serafettin
  Tasci}, \bibinfo{person}{Shalini Ghosh}, \bibinfo{person}{Jingwen Zhu},
  \bibinfo{person}{Junting Zhang}, {and} \bibinfo{person}{Larry Heck}.}
  \bibinfo{year}{2019}\natexlab{}.
\newblock \showarticletitle{RILOD: Near real-time incremental learning for
  object detection at the edge}. In \bibinfo{booktitle}{\emph{Proceedings of
  the 4th ACM/IEEE Symposium on Edge Computing}}. \bibinfo{pages}{113--126}.
\newblock


\bibitem[Li et~al\mbox{.}(2018)]%
        {mobi}
\bibfield{author}{\bibinfo{person}{Yiting Li}, \bibinfo{person}{Haisong Huang},
  \bibinfo{person}{Qingsheng Xie}, \bibinfo{person}{Liguo Yao}, {and}
  \bibinfo{person}{Qipeng Chen}.} \bibinfo{year}{2018}\natexlab{}.
\newblock \showarticletitle{Research on a surface defect detection algorithm
  based on MobileNet-SSD}.
\newblock \bibinfo{journal}{\emph{Applied Sciences}} \bibinfo{volume}{8},
  \bibinfo{number}{9} (\bibinfo{year}{2018}), \bibinfo{pages}{1678}.
\newblock


\bibitem[Li et~al\mbox{.}(2020)]%
        {reducto}
\bibfield{author}{\bibinfo{person}{Yuanqi Li}, \bibinfo{person}{Arthi
  Padmanabhan}, \bibinfo{person}{Pengzhan Zhao}, \bibinfo{person}{Yufei Wang},
  \bibinfo{person}{Guoqing~Harry Xu}, {and} \bibinfo{person}{Ravi Netravali}.}
  \bibinfo{year}{2020}\natexlab{}.
\newblock \showarticletitle{Reducto: On-camera filtering for resource-efficient
  real-time video analytics}. In \bibinfo{booktitle}{\emph{Proceedings of the
  Annual conference of the ACM Special Interest Group on Data Communication on
  the applications, technologies, architectures, and protocols for computer
  communication}}. \bibinfo{pages}{359--376}.
\newblock


\bibitem[Lin et~al\mbox{.}(2021)]%
        {a7}
\bibfield{author}{\bibinfo{person}{Jie Lin}, \bibinfo{person}{Peng Yang},
  \bibinfo{person}{Wen Wu}, \bibinfo{person}{Ning Zhang}, \bibinfo{person}{Tao
  Han}, {and} \bibinfo{person}{Li Yu}.} \bibinfo{year}{2021}\natexlab{}.
\newblock \showarticletitle{Edge learning for low-latency video analytics:
  Query scheduling and resource allocation}. In \bibinfo{booktitle}{\emph{2021
  IEEE 18th International Conference on Mobile Ad Hoc and Smart Systems
  (MASS)}}. IEEE, \bibinfo{pages}{252--259}.
\newblock


\bibitem[Lin et~al\mbox{.}(2022)]%
        {a4}
\bibfield{author}{\bibinfo{person}{Jie Lin}, \bibinfo{person}{Peng Yang},
  \bibinfo{person}{Ning Zhang}, \bibinfo{person}{Feng Lyu},
  \bibinfo{person}{Xianfu Chen}, {and} \bibinfo{person}{Li Yu}.}
  \bibinfo{year}{2022}\natexlab{}.
\newblock \showarticletitle{Low-latency edge video analytics for on-road
  perception of autonomous ground vehicles}.
\newblock \bibinfo{journal}{\emph{IEEE Transactions on Industrial Informatics}}
  \bibinfo{volume}{19}, \bibinfo{number}{2} (\bibinfo{year}{2022}),
  \bibinfo{pages}{1512--1523}.
\newblock


\bibitem[Lin et~al\mbox{.}(2014)]%
        {c4}
\bibfield{author}{\bibinfo{person}{Tsung-Yi Lin}, \bibinfo{person}{Michael
  Maire}, \bibinfo{person}{Serge Belongie}, \bibinfo{person}{James Hays},
  \bibinfo{person}{Pietro Perona}, \bibinfo{person}{Deva Ramanan},
  \bibinfo{person}{Piotr Doll{\'a}r}, {and} \bibinfo{person}{C~Lawrence
  Zitnick}.} \bibinfo{year}{2014}\natexlab{}.
\newblock \showarticletitle{Microsoft coco: Common objects in context}. In
  \bibinfo{booktitle}{\emph{Computer Vision--ECCV 2014: 13th European
  Conference, Zurich, Switzerland, September 6-12, 2014, Proceedings, Part V
  13}}. Springer, \bibinfo{pages}{740--755}.
\newblock


\bibitem[Liu et~al\mbox{.}(2022)]%
        {acm2}
\bibfield{author}{\bibinfo{person}{Shengzhong Liu}, \bibinfo{person}{Tianshi
  Wang}, \bibinfo{person}{Jinyang Li}, \bibinfo{person}{Dachun Sun},
  \bibinfo{person}{Mani Srivastava}, {and} \bibinfo{person}{Tarek Abdelzaher}.}
  \bibinfo{year}{2022}\natexlab{}.
\newblock \showarticletitle{Adamask: Enabling machine-centric video streaming
  with adaptive frame masking for dnn inference offloading}. In
  \bibinfo{booktitle}{\emph{Proceedings of the 30th ACM international
  conference on multimedia}}. \bibinfo{pages}{3035--3044}.
\newblock


\bibitem[Maltoni and Lomonaco(2019)]%
        {d1}
\bibfield{author}{\bibinfo{person}{Davide Maltoni} {and}
  \bibinfo{person}{Vincenzo Lomonaco}.} \bibinfo{year}{2019}\natexlab{}.
\newblock \showarticletitle{Continuous learning in single-incremental-task
  scenarios}.
\newblock \bibinfo{journal}{\emph{Neural Networks}}  \bibinfo{volume}{116}
  (\bibinfo{year}{2019}), \bibinfo{pages}{56--73}.
\newblock


\bibitem[Mehta and Ozturk(2018)]%
        {obj200}
\bibfield{author}{\bibinfo{person}{Rakesh Mehta} {and}
  \bibinfo{person}{Cemalettin Ozturk}.} \bibinfo{year}{2018}\natexlab{}.
\newblock \showarticletitle{Object detection at 200 frames per second}. In
  \bibinfo{booktitle}{\emph{Proceedings of the European Conference on Computer
  Vision (ECCV) Workshops}}. \bibinfo{pages}{0--0}.
\newblock


\bibitem[Metropolis et~al\mbox{.}(1953)]%
        {metro}
\bibfield{author}{\bibinfo{person}{Nicholas Metropolis},
  \bibinfo{person}{Arianna~W Rosenbluth}, \bibinfo{person}{Marshall~N
  Rosenbluth}, \bibinfo{person}{Augusta~H Teller}, {and}
  \bibinfo{person}{Edward Teller}.} \bibinfo{year}{1953}\natexlab{}.
\newblock \showarticletitle{Equation of state calculations by fast computing
  machines}.
\newblock \bibinfo{journal}{\emph{The journal of chemical physics}}
  \bibinfo{volume}{21}, \bibinfo{number}{6} (\bibinfo{year}{1953}),
  \bibinfo{pages}{1087--1092}.
\newblock


\bibitem[Mullapudi et~al\mbox{.}(2019)]%
        {JIT}
\bibfield{author}{\bibinfo{person}{Ravi~Teja Mullapudi},
  \bibinfo{person}{Steven Chen}, \bibinfo{person}{Keyi Zhang},
  \bibinfo{person}{Deva Ramanan}, {and} \bibinfo{person}{Kayvon Fatahalian}.}
  \bibinfo{year}{2019}\natexlab{}.
\newblock \showarticletitle{Online model distillation for efficient video
  inference}. In \bibinfo{booktitle}{\emph{Proceedings of the IEEE/CVF
  International conference on computer vision}}. \bibinfo{pages}{3573--3582}.
\newblock


\bibitem[Murad et~al\mbox{.}(2022)]%
        {acm1}
\bibfield{author}{\bibinfo{person}{Taslim Murad}, \bibinfo{person}{Anh Nguyen},
  {and} \bibinfo{person}{Zhisheng Yan}.} \bibinfo{year}{2022}\natexlab{}.
\newblock \showarticletitle{DAO: Dynamic Adaptive Offloading for Video
  Analytics}. In \bibinfo{booktitle}{\emph{Proceedings of the 30th ACM
  International Conference on Multimedia}}. \bibinfo{pages}{3017--3025}.
\newblock


\bibitem[Noghabi et~al\mbox{.}(2020)]%
        {emerging}
\bibfield{author}{\bibinfo{person}{Shadi~A Noghabi}, \bibinfo{person}{Landon
  Cox}, \bibinfo{person}{Sharad Agarwal}, {and} \bibinfo{person}{Ganesh
  Ananthanarayanan}.} \bibinfo{year}{2020}\natexlab{}.
\newblock \showarticletitle{The emerging landscape of edge computing}.
\newblock \bibinfo{journal}{\emph{GetMobile: Mobile Computing and
  Communications}} \bibinfo{volume}{23}, \bibinfo{number}{4}
  (\bibinfo{year}{2020}), \bibinfo{pages}{11--20}.
\newblock


\bibitem[NVIDIA Corporation.(2017)]%
        {tx2}
NVIDIA Corporation. \bibinfo{year}{2017}\natexlab{}.
\newblock \bibinfo{booktitle}{\emph{NVIDIA Jetson TX2 Developer Kit.}}
\newblock
\urldef\tempurl%
\url{https://www.nvidia.com/en-us/autonomousmachines/embedded-systems/jetson-tx2.}
\showURL{%
Retrieved 2017 from \tempurl}


\bibitem[NVIDIA Corporation.(2019)]%
        {nano}
NVIDIA Corporation. \bibinfo{year}{2019}\natexlab{}.
\newblock \bibinfo{booktitle}{\emph{NVIDIA Jetson Nano Developer Kit.}}
\newblock
\urldef\tempurl%
\url{https://developer.nvidia.com/embedded/jetson-nano-developer-kit.}
\showURL{%
Retrieved 2019 from \tempurl}


\bibitem[Olatunji and Cheng(2019)]%
        {survey}
\bibfield{author}{\bibinfo{person}{Iyiola~E Olatunji} {and}
  \bibinfo{person}{Chun-Hung Cheng}.} \bibinfo{year}{2019}\natexlab{}.
\newblock \showarticletitle{Video analytics for visual surveillance and
  applications: An overview and survey}.
\newblock \bibinfo{journal}{\emph{Machine Learning Paradigms: Applications of
  Learning and Analytics in Intelligent Systems}} (\bibinfo{year}{2019}),
  \bibinfo{pages}{475--515}.
\newblock


\bibitem[Paul et~al\mbox{.}(2021)]%
        {aqua}
\bibfield{author}{\bibinfo{person}{Sibendu Paul}, \bibinfo{person}{Utsav
  Drolia}, \bibinfo{person}{Y~Charlie Hu}, {and} \bibinfo{person}{Srimat~T
  Chakradhar}.} \bibinfo{year}{2021}\natexlab{}.
\newblock \showarticletitle{Aqua: Analytical quality assessment for optimizing
  video analytics systems}. In \bibinfo{booktitle}{\emph{2021 IEEE/ACM
  Symposium on Edge Computing (SEC)}}. IEEE, \bibinfo{pages}{135--147}.
\newblock


\bibitem[Qi et~al\mbox{.}(2020)]%
        {temp}
\bibfield{author}{\bibinfo{person}{Zhaobo Qi}, \bibinfo{person}{Shuhui Wang},
  \bibinfo{person}{Chi Su}, \bibinfo{person}{Li Su}, \bibinfo{person}{Weigang
  Zhang}, {and} \bibinfo{person}{Qingming Huang}.}
  \bibinfo{year}{2020}\natexlab{}.
\newblock \showarticletitle{Modeling temporal concept receptive field
  dynamically for untrimmed video analysis}. In
  \bibinfo{booktitle}{\emph{Proceedings of the 28th ACM International
  Conference on Multimedia}}. \bibinfo{pages}{3798--3806}.
\newblock


\bibitem[Ran et~al\mbox{.}(2018)]%
        {deep}
\bibfield{author}{\bibinfo{person}{Xukan Ran}, \bibinfo{person}{Haolianz Chen},
  \bibinfo{person}{Xiaodan Zhu}, \bibinfo{person}{Zhenming Liu}, {and}
  \bibinfo{person}{Jiasi Chen}.} \bibinfo{year}{2018}\natexlab{}.
\newblock \showarticletitle{Deepdecision: A mobile deep learning framework for
  edge video analytics}. In \bibinfo{booktitle}{\emph{IEEE INFOCOM 2018-IEEE
  conference on computer communications}}. IEEE, \bibinfo{pages}{1421--1429}.
\newblock


\bibitem[Rivas et~al\mbox{.}(2022)]%
        {spe}
\bibfield{author}{\bibinfo{person}{Daniel Rivas}, \bibinfo{person}{Francesc
  Guim}, \bibinfo{person}{Jord{\`a} Polo}, \bibinfo{person}{Pubudu~M Silva},
  \bibinfo{person}{Josep~Ll Berral}, {and} \bibinfo{person}{David Carrera}.}
  \bibinfo{year}{2022}\natexlab{}.
\newblock \showarticletitle{Towards automatic model specialization for edge
  video analytics}.
\newblock \bibinfo{journal}{\emph{Future Generation Computer Systems}}
  \bibinfo{volume}{134} (\bibinfo{year}{2022}), \bibinfo{pages}{399--413}.
\newblock


\bibitem[Singla(2014)]%
        {framediff}
\bibfield{author}{\bibinfo{person}{Nishu Singla}.}
  \bibinfo{year}{2014}\natexlab{}.
\newblock \showarticletitle{Motion detection based on frame difference method}.
\newblock \bibinfo{journal}{\emph{International Journal of Information \&
  Computation Technology}} \bibinfo{volume}{4}, \bibinfo{number}{15}
  (\bibinfo{year}{2014}), \bibinfo{pages}{1559--1565}.
\newblock


\bibitem[Tan and Le(2019)]%
        {effi}
\bibfield{author}{\bibinfo{person}{Mingxing Tan} {and} \bibinfo{person}{Quoc
  Le}.} \bibinfo{year}{2019}\natexlab{}.
\newblock \showarticletitle{Efficientnet: Rethinking model scaling for
  convolutional neural networks}. In \bibinfo{booktitle}{\emph{International
  conference on machine learning}}. PMLR, \bibinfo{pages}{6105--6114}.
\newblock


\bibitem[TensorRT(2018)]%
        {c6}
TensorRT \bibinfo{year}{2018}\natexlab{}.
\newblock \bibinfo{booktitle}{\emph{TensorRT: Inference Accelerator for
  Precision Inference at Scale.}}
\newblock
\urldef\tempurl%
\url{https://developer.nvidia.com/tensorrt}
\showURL{%
Retrieved 2018 from \tempurl}


\bibitem[Wang et~al\mbox{.}(2022)]%
        {a5}
\bibfield{author}{\bibinfo{person}{Chengzhi Wang}, \bibinfo{person}{Peng Yang},
  \bibinfo{person}{Jie Lin}, \bibinfo{person}{Wen Wu}, {and}
  \bibinfo{person}{Ning Zhang}.} \bibinfo{year}{2022}\natexlab{}.
\newblock \showarticletitle{Object-Based Resolution Selection for Efficient
  Edge-Assisted Multi-Task Video Analytics}. In
  \bibinfo{booktitle}{\emph{GLOBECOM 2022-2022 IEEE Global Communications
  Conference}}. IEEE, \bibinfo{pages}{5081--5086}.
\newblock


\bibitem[Wang et~al\mbox{.}(2020)]%
        {jcab}
\bibfield{author}{\bibinfo{person}{Can Wang}, \bibinfo{person}{Sheng Zhang},
  \bibinfo{person}{Yu Chen}, \bibinfo{person}{Zhuzhong Qian},
  \bibinfo{person}{Jie Wu}, {and} \bibinfo{person}{Mingjun Xiao}.}
  \bibinfo{year}{2020}\natexlab{}.
\newblock \showarticletitle{Joint configuration adaptation and bandwidth
  allocation for edge-based real-time video analytics}. In
  \bibinfo{booktitle}{\emph{IEEE INFOCOM 2020-IEEE Conference on Computer
  Communications}}. IEEE, \bibinfo{pages}{257--266}.
\newblock


\bibitem[Wang and Yoon(2021)]%
        {b7}
\bibfield{author}{\bibinfo{person}{Lin Wang} {and} \bibinfo{person}{Kuk-Jin
  Yoon}.} \bibinfo{year}{2021}\natexlab{}.
\newblock \showarticletitle{Knowledge distillation and student-teacher learning
  for visual intelligence: A review and new outlooks}.
\newblock \bibinfo{journal}{\emph{IEEE transactions on pattern analysis and
  machine intelligence}} \bibinfo{volume}{44}, \bibinfo{number}{6}
  (\bibinfo{year}{2021}), \bibinfo{pages}{3048--3068}.
\newblock


\bibitem[Wang et~al\mbox{.}(2019a)]%
        {drl}
\bibfield{author}{\bibinfo{person}{Shangguang Wang}, \bibinfo{person}{Yan Guo},
  \bibinfo{person}{Ning Zhang}, \bibinfo{person}{Peng Yang},
  \bibinfo{person}{Ao Zhou}, {and} \bibinfo{person}{Xuemin Shen}.}
  \bibinfo{year}{2019}\natexlab{a}.
\newblock \showarticletitle{Delay-aware microservice coordination in mobile
  edge computing: A reinforcement learning approach}.
\newblock \bibinfo{journal}{\emph{IEEE Transactions on Mobile Computing}}
  \bibinfo{volume}{20}, \bibinfo{number}{3} (\bibinfo{year}{2019}),
  \bibinfo{pages}{939--951}.
\newblock


\bibitem[Wang et~al\mbox{.}(2019b)]%
        {offload}
\bibfield{author}{\bibinfo{person}{Yue Wang}, \bibinfo{person}{Xiaofeng Tao},
  \bibinfo{person}{Xuefei Zhang}, \bibinfo{person}{Ping Zhang}, {and}
  \bibinfo{person}{Y~Thomas Hou}.} \bibinfo{year}{2019}\natexlab{b}.
\newblock \showarticletitle{Cooperative task offloading in three-tier mobile
  computing networks: An ADMM framework}.
\newblock \bibinfo{journal}{\emph{IEEE Transactions on Vehicular Technology}}
  \bibinfo{volume}{68}, \bibinfo{number}{3} (\bibinfo{year}{2019}),
  \bibinfo{pages}{2763--2776}.
\newblock


\bibitem[Wu et~al\mbox{.}(2020)]%
        {a6}
\bibfield{author}{\bibinfo{person}{Wen Wu}, \bibinfo{person}{Peng Yang},
  \bibinfo{person}{Weiting Zhang}, \bibinfo{person}{Conghao Zhou}, {and}
  \bibinfo{person}{Xuemin Shen}.} \bibinfo{year}{2020}\natexlab{}.
\newblock \showarticletitle{Accuracy-guaranteed collaborative DNN inference in
  industrial IoT via deep reinforcement learning}.
\newblock \bibinfo{journal}{\emph{IEEE Transactions on Industrial Informatics}}
  \bibinfo{volume}{17}, \bibinfo{number}{7} (\bibinfo{year}{2020}),
  \bibinfo{pages}{4988--4998}.
\newblock


\bibitem[Yang et~al\mbox{.}(2023)]%
        {dt}
\bibfield{author}{\bibinfo{person}{Peng Yang}, \bibinfo{person}{Jiawei Hou},
  \bibinfo{person}{Li Yu}, \bibinfo{person}{Wenxiong Chen}, {and}
  \bibinfo{person}{Ye Wu}.} \bibinfo{year}{2023}\natexlab{}.
\newblock \showarticletitle{Edge-coordinated energy-efficient video analytics
  for digital twin in 6G}.
\newblock \bibinfo{journal}{\emph{China Communications}} \bibinfo{volume}{20},
  \bibinfo{number}{2} (\bibinfo{year}{2023}), \bibinfo{pages}{14--25}.
\newblock


\bibitem[Yang et~al\mbox{.}(2019)]%
        {a2}
\bibfield{author}{\bibinfo{person}{Peng Yang}, \bibinfo{person}{Feng Lyu},
  \bibinfo{person}{Wen Wu}, \bibinfo{person}{Ning Zhang}, \bibinfo{person}{Li
  Yu}, {and} \bibinfo{person}{Xuemin~Sherman Shen}.}
  \bibinfo{year}{2019}\natexlab{}.
\newblock \showarticletitle{Edge coordinated query configuration for
  low-latency and accurate video analytics}.
\newblock \bibinfo{journal}{\emph{IEEE Transactions on Industrial Informatics}}
  \bibinfo{volume}{16}, \bibinfo{number}{7} (\bibinfo{year}{2019}),
  \bibinfo{pages}{4855--4864}.
\newblock


\bibitem[Yolov5 v6.2(2022)]%
        {c1}
Yolov5 v6.2 \bibinfo{year}{2022}\natexlab{}.
\newblock \bibinfo{booktitle}{\emph{YOLOv5 by Ultralytics}}.
\newblock
\urldef\tempurl%
\url{https://github.com/ultralytics/yolov5}
\showURL{%
Retrieved Feb, 2022 from \tempurl}


\bibitem[Zeng et~al\mbox{.}(2020)]%
        {distream}
\bibfield{author}{\bibinfo{person}{Xiao Zeng}, \bibinfo{person}{Biyi Fang},
  \bibinfo{person}{Haichen Shen}, {and} \bibinfo{person}{Mi Zhang}.}
  \bibinfo{year}{2020}\natexlab{}.
\newblock \showarticletitle{Distream: scaling live video analytics with
  workload-adaptive distributed edge intelligence}. In
  \bibinfo{booktitle}{\emph{Proceedings of the 18th Conference on Embedded
  Networked Sensor Systems}}. \bibinfo{pages}{409--421}.
\newblock


\bibitem[Zhang et~al\mbox{.}(2018a)]%
        {aws}
\bibfield{author}{\bibinfo{person}{Ben Zhang}, \bibinfo{person}{Xin Jin},
  \bibinfo{person}{Sylvia Ratnasamy}, \bibinfo{person}{John Wawrzynek}, {and}
  \bibinfo{person}{Edward~A Lee}.} \bibinfo{year}{2018}\natexlab{a}.
\newblock \showarticletitle{Awstream: Adaptive wide-area streaming analytics}.
  In \bibinfo{booktitle}{\emph{Proceedings of the 2018 Conference of the ACM
  Special Interest Group on Data Communication}}. \bibinfo{pages}{236--252}.
\newblock


\bibitem[Zhang et~al\mbox{.}(2022)]%
        {casva}
\bibfield{author}{\bibinfo{person}{Miao Zhang}, \bibinfo{person}{Fangxin Wang},
  {and} \bibinfo{person}{Jiangchuan Liu}.} \bibinfo{year}{2022}\natexlab{}.
\newblock \showarticletitle{Casva: Configuration-adaptive streaming for live
  video analytics}. In \bibinfo{booktitle}{\emph{IEEE INFOCOM 2022-IEEE
  Conference on Computer Communications}}. IEEE, \bibinfo{pages}{2168--2177}.
\newblock


\bibitem[Zhang et~al\mbox{.}(2021)]%
        {elf}
\bibfield{author}{\bibinfo{person}{Wuyang Zhang}, \bibinfo{person}{Zhezhi He},
  \bibinfo{person}{Luyang Liu}, \bibinfo{person}{Zhenhua Jia},
  \bibinfo{person}{Yunxin Liu}, \bibinfo{person}{Marco Gruteser},
  \bibinfo{person}{Dipankar Raychaudhuri}, {and} \bibinfo{person}{Yanyong
  Zhang}.} \bibinfo{year}{2021}\natexlab{}.
\newblock \showarticletitle{Elf: accelerate high-resolution mobile deep vision
  with content-aware parallel offloading}. In
  \bibinfo{booktitle}{\emph{Proceedings of the 27th Annual International
  Conference on Mobile Computing and Networking}}. \bibinfo{pages}{201--214}.
\newblock


\bibitem[Zhang et~al\mbox{.}(2018b)]%
        {b2}
\bibfield{author}{\bibinfo{person}{Xiangyu Zhang}, \bibinfo{person}{Xinyu
  Zhou}, \bibinfo{person}{Mengxiao Lin}, {and} \bibinfo{person}{Jian Sun}.}
  \bibinfo{year}{2018}\natexlab{b}.
\newblock \showarticletitle{Shufflenet: An extremely efficient convolutional
  neural network for mobile devices}. In \bibinfo{booktitle}{\emph{Proceedings
  of the IEEE conference on computer vision and pattern recognition}}.
  \bibinfo{pages}{6848--6856}.
\newblock


\end{thebibliography}

\end{document}